\documentclass[final]{siamltex}

\usepackage{epsfig}

\title{Dynamics of point Josephson junctions in a microstrip line}
\author{J.-G.~Caputo\footnote{Laboratoire de Math\'ematiques, 
INSA de Rouen, B.P. 8, Avenue de l'Universite
76801 Saint-Etienne du Rouvray, France.
E-mail: caputo@insa-rouen.fr, loukitch@insa-rouen.fr}
 and L. Loukitch$^*$}

\begin{document}

\maketitle


%
\begin{abstract}
We analyze a new long wave model describing the electrodynamics of
an array of point Josephson junctions in a superconducting cavity. It consists
in a wave equation with Dirac delta function sine nonlinearities. 
We introduce an adapted spectral problem whose spectrum gives the
resonances in the current-voltage characteristic curve of any array.
Using the associated inner product and eigenmodes, we establish that
at the resonances the solution is described by two simple ordinary
differential equations.
\end{abstract}

\begin{keywords}
Josephson junctions, Dirac distribution, sine Gordon , 
current voltage characteristic
spectral problem, resonance
\end{keywords}

\begin{AMS}
35Qxx, 46Fxx, 35Pxx 
\end{AMS}

\pagestyle{myheadings}
\thispagestyle{plain}
\markboth{J.G. Caputo and L. Loukitch}{Dynamics of point Josephson junctions 
in a microstrip.}

\section{Introduction}
\label{sec:intro}


The macroscopic state of a superconductor is described
by a complex field the order parameter. For low T$_c$ superconductors
it can be assumed that only the phase of the order parameter
varies. The coupling of two such superconductors
across a thin oxide layer is described
by the Josephson equations \cite{josephson}.
\begin{equation}\label{josephson}
V=\frac{d\Phi}{dt},~~~I=sJ_c\sin({\Phi \over \Phi_0})~,
\end{equation}
where $\Phi$ is the phase difference between the top and bottom
superconductor, $V$ and $I$ are, respectively, the voltage and current across
the barrier, $s$ is the contact surface, $J_c$ is the critical current
density and $\Phi_0=\hbar/2e$ is the reduced flux quantum.
These two Josephson relations together with Maxwell's equations imply
the modulation of DC current by an external magnetic field in the static
regime and the conversion of AC current into microwave radiation
\cite{Barone,Likharev}. Such Josephson junctions are then unique electronic
systems for applications like the detection of magnetic fields, ultra
fast electronics\cite{Likharev} and microwave sources and signal mixers
\cite{Salez},\cite{bsf09}.

For the applications the devices are often associated to form
arrays. The junctions can be in parallel or in series.
The series arrays can lead to synchronization\cite{vbsl02}.
and deliver more output power for some applications. Their
description is however more complex and we will not consider it
here. Parallel arrays where the junctions are embedded
between two superconducting planes are now relatively easy to prepare and the
junction is protected from the atmosphere. In addition one
can easily prepare an array with junctions of specific
sizes and positions. Such non uniform arrays have been produced
and analyzed in particular by Salez and co-workers at the Observatory of
Paris. For these systems, the phase difference $\Phi$
satisfies an inhomogeneous 2D damped driven sine Gordon equation \cite{cfv95}
resulting from Maxwell's equations and the Josephson constitutive relations
(\ref{josephson}). The damping is due to the normal electrons and the driving
through the boundary conditions with an external current or
magnetic field applied to the device.

To model such arrays authors have used lumped models where
the spatial dependence between the junctions is omitted. This
obliterates the wave features of the solution and does not
describe well the experiments. Solving numerically the full
two dimensional problem is of course possible, however it
does not lead to understand simply the role of the parameters.
Similar difficulties occur with global (hard) analysis. Consider
for example the problem of finding the maximum current giving
a static solution for a given magnetic field. 
Using such global analysis we obtained bounds \cite{cftv03} on the
gradient of the solution that were independent of the area of the junctions
so that little information could be obtained from them.
To overcome these difficulties we recently introduced a continuous/discrete
model that preserves the
continuity of the phase and its normal gradient across the junction interface
and where the phase is assumed constant in the junctions. The relative
simplicity of the model allowed an unprecedented understanding of the
static problem\cite{cl06} and gave excellent agreement with the
complex static response of the array\cite{sblc07}. Additionally the model allows
to solve the inverse problem of building a device that produces a given static
behavior\cite{cl08}.

The dynamic behavior of Josephson junctions is characterized by
the current-voltage (I-V) characteristic curve.
To understand this one needs to analyze periodic solutions of the problem.
For a homogeneous long junction Kulik\cite{kulik} developed
a formalism using a high voltage ansatz and obtained average equations
describing the I-V curve.
This approach was extended by Cirillo et al \cite{cgssv98} who showed that
a magnetic field $\Gamma$ reinforces the cavity modes such that 
$\Gamma=n \pi/l$. Using this approach these authors obtained excellent 
quantitative agreement with their experimental results.
For arrays of equidistant junctions, 
a recent study by Pfeiffer et 
al\cite{pasu08} analyzed the fine features of the first resonant step 
in terms of Cerenkov radiation between a sine-Gordon discrete kink (fluxon)
and a cavity mode.
For one junction in a cavity, our theoretical
study \cite{cl05} 
revealed that the junction could stop waves across the cavity
or enhance them throughout. We also found kink like solutions for the
problem \cite{cl07} and explained some features of the current-voltage
characteristics. Here we analyze theoretically and
numerically the model in particular
when there is a capacity miss-match between the junctions and the cavity.
This capacity ratio is usually large in experiments because the
oxide layer in the junction is about 10 Angstroms while it is about
0.2 Micron in the strip. Taking this miss-match into account
we introduce an associated linear problem which
enables us to predict the position of the resonances in the current-voltage
curve for any array. This linear problem defines eigenvalues and eigenmodes
orthogonal with respect to an inner product that we establish.
At these resonances the solution just contains the Goldstone mode
and the corresponding eigenmode so the dynamics is described by two
simple amplitude equations that we present and analyze. \\
The article is organized as such. After introducing the model in section
2, we analyze it in section 3, establish the periodicity of the current
voltage curve as a function of the magnetic field and simplify the
model using the time averaged (high voltage) solution. The resonances
are studied in section 4 where we define the appropriate spectral problem
for any given array
and find its spectrum and associated inner product. Using the latter
we analyze numerically the current voltage curves in section 5. In particular
we establish and analyze the amplitude equations describing the system
at resonance and discuss the situation for arrays containing many junctions.

\section{The model}
\subsection{The $2D$ problem}
\begin{figure}
\centerline{\epsfig{file=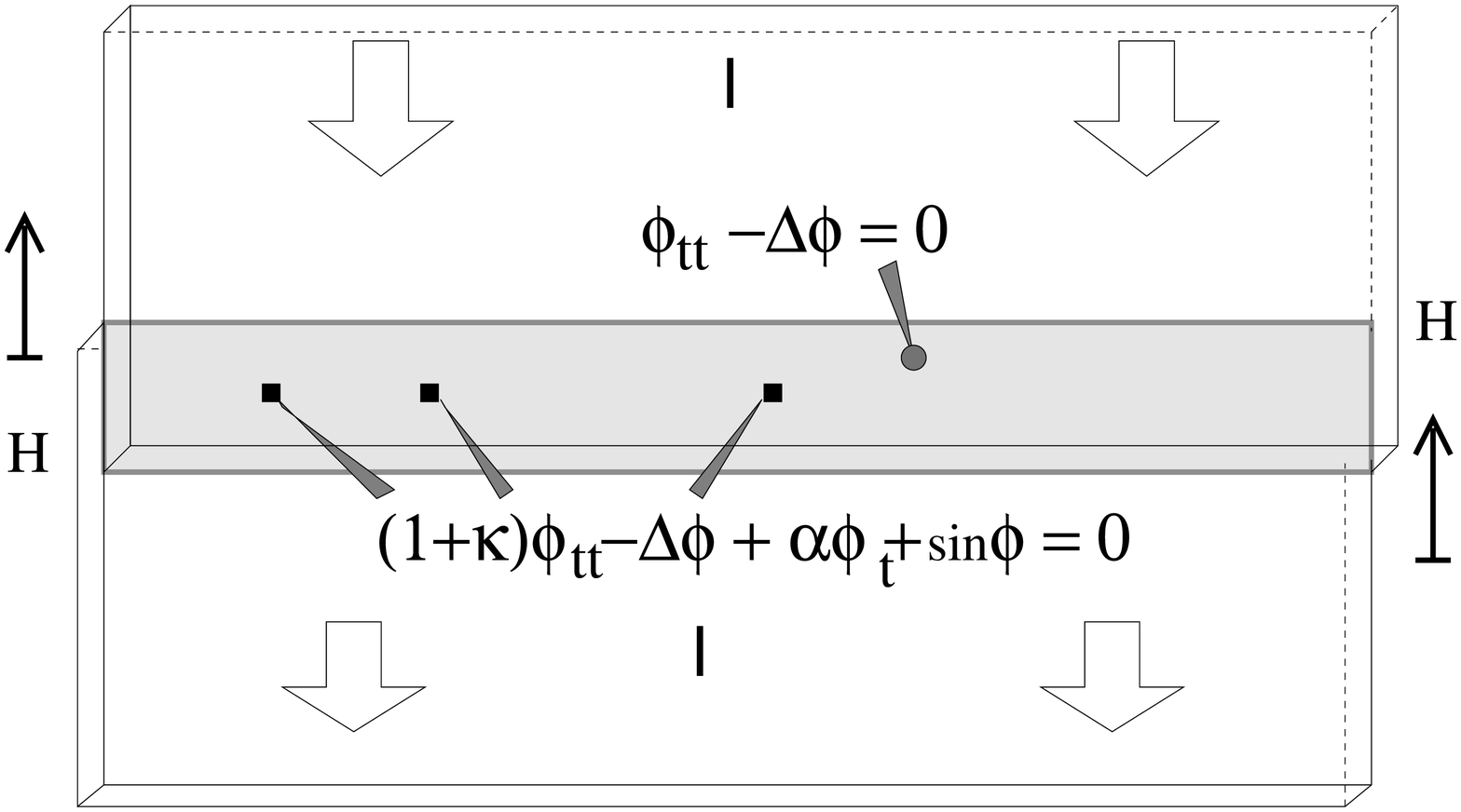,width=0.49\linewidth,angle=0}
\epsfig{file=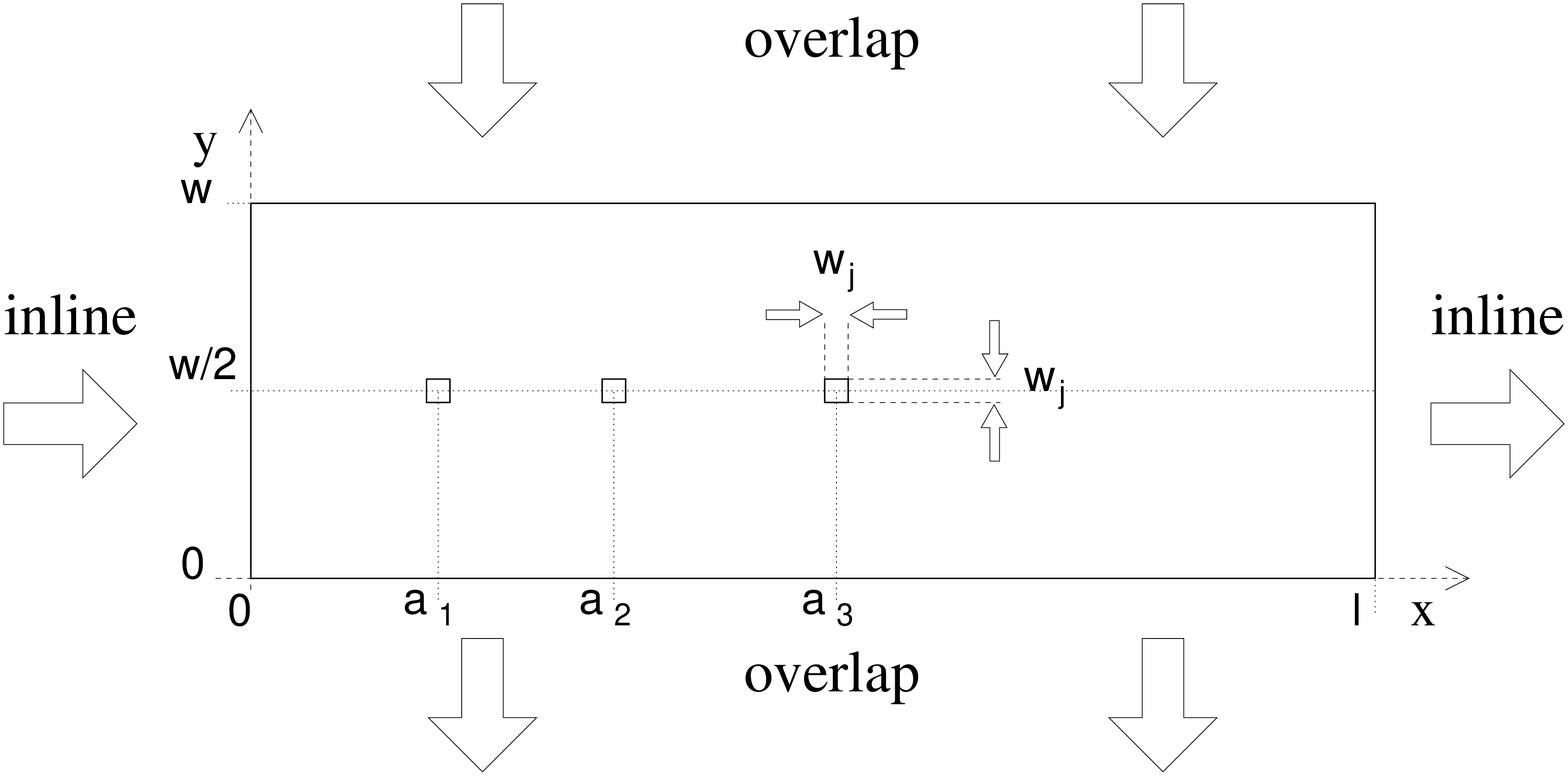,width=0.49\linewidth,angle=0}}
\caption{The left panel shows the top view of a superconducting
microstrip line containing three Josephson junctions.
The parameters $H,I$ and $\phi$ are respectively the applied 
magnetic field, the current and the
phase difference between the two superconducting layers.
The right panel
shows the associated 2D domain of size $l \times w$ containing $n=3$ junctions
placed at the positions $y=w/2$ and $x=a_i,~i=1,n$.}
\label{f1}
\end{figure}

The device we
model (see Fig.\ref{f1}) is composed of two
overlapping superconducting layers, in which stand
small\footnote{compared to the Josephson coherence length $\lambda_J$}
Josephson junctions.
 Using the Josephson constitutive equations
(eq.\ref{josephson}) and Maxwell's equation, one obtains
the following inhomogeneous sine-Gordon equation for $\Phi$ in the
junction region $\Omega_j$ \cite{Barone,Likharev}
\begin{equation}\label{2D_sg}
C_j \Phi_{tt} - \frac{1}{L} \Delta \Phi + J_c
\sin\left(\frac{\Phi}{\Phi_0}\right)+\frac{1}{R} \Phi_t=0\;,
\end{equation}
where $C_j$ is the capacity of the junction 
per unit area, $R$ the resistance per unit
area due to normal electrons. The branch inductance $L =\mu_0 d_0$ involves the
magnetic thickness $d_0$, a quantity used by experimentalists. 
In the microstrip $\Omega-\Omega_j$, the Josephson and 
quasi-particle currents are
absent so one obtains the wave equation,
\begin{equation}\label{2D_onde}
C_l \Phi_{tt} - \frac{1}{L_l} \Delta \Phi = 0~~,
\end{equation}
where the $l$ subscripts indicate that we are in the linear region
(i.e.: outside the junction). 
For inhomogeneous circuits like the one of Fig. \ref{f1} these 
two equations can be written as \cite{cl05,louk_th}
\begin{equation}\label{dimsgi}
{C_l} \Phi_{tt} -\nabla \left({1 \over L(x)}\nabla \Phi \right)
+ g(x,y)\left [ (C_j- C_l) \Phi_{tt}
+ J_c \sin \left({\Phi \over \Phi_0}\right) + {\Phi_t \over R} \right ] =0,\end{equation}
where $g=1$ in the junctions and 0 outside. 
This formulation guarantees the continuity of the phase and its normal gradient
across the interfaces.
At this point we will assume the same surface inductance $L$ in the junctions
and linear region. This simplifies greatly the formulation and can be realized
in practical situations.
To normalize the equation, we introduce the units of length and time,
respectively the Josephson length $\lambda_J$ and plasma frequency $\omega_p$
\begin{equation}\label{ljop}
\lambda_J = \sqrt{\frac{\Phi_0}{J_c L_l}}\;,~~~\frac{1}{\omega_p}=
\sqrt{\frac{C_l \Phi_0}{J_c}}\;.
\end{equation}
Finally we normalize space, time and phase as
\begin{equation}\label{norma}
\tilde{x}=x/\lambda_J,~~
\tilde{y}=y/\lambda_J,~~\tilde{t}=t \omega_p,
~~\varphi = \Phi / \Phi_0,
\end{equation}
to get the normalized 2D inhomogeneous perturbed sine-Gordon equation
\begin{equation}\label{sg2d}
\varphi_{\tilde t \tilde t} -\Delta \varphi + g( \tilde x, \tilde y)(\kappa \varphi_{ \tilde t \tilde t} +
\alpha \varphi_{ \tilde t} + \sin\varphi  )= 0\end{equation}
where the coefficients $\alpha$
and $\kappa$ are
\begin{equation}\label{alphakappa}
\kappa = \frac{{C_j}}{{C_l}}-1,~~~\alpha=\frac{1}{{R}}
\sqrt{\frac{\Phi_0}{J_c {C_l}}}\;.
\end{equation}

The boundary conditions are
\begin{eqnarray} \label{bc2d}
\varphi_{\tilde x}|_{({\tilde x}=0)} = {\tilde H} + (1-\nu) \frac{\tilde I}{2\tilde w},~~~
\varphi_{\tilde x}|_{({\tilde x}={\tilde l})} = {\tilde H} - (1-\nu)
\frac{\tilde I}{2\tilde w}, \\
\varphi_{\tilde y}|_{({\tilde y}=0)} =  \nu \frac{\tilde I}{2\tilde l},~~~
\varphi_{\tilde y}|_{({\tilde y}={\tilde w})} =  -\nu \frac{\tilde I}{2\tilde l},
\end{eqnarray}
where
\begin{equation}\label{dimH_I}
\tilde H = H \frac{\lambda_J d_0}{\Phi_0}\;,~~
\tilde I = I \frac{\mu_0 d_0}{\Phi_0} =  \frac{L I}{\Phi_0}\;,
\end{equation}
and $l$ and $w$ are normalized by $\lambda_J$.
After this section all tildes will be omitted for simplicity.

\subsection{The $1D$ model}

This equation is difficult to analyze and its solutions can only be obtained
numerically. In addition most real devices have a width that is much smaller
than their length and the junctions are distributed symmetrically so that it
is reasonable to reduce the problem to one dimension. To do this we
expand $\varphi$ on transverse Fourier modes
\begin{equation}\label{2Dto1D}
\varphi(x,y,t) = \frac{\nu \gamma}{2 l}\left(y-\frac{w}{2}\right)^2
+ \sum^{\infty}_{n=0} \phi_n(x,t) \cos \left(\frac{n \pi y}{w}\right)\;,
\end{equation}
where $\gamma=I/w$ and 
the first term takes care of the boundary condition (\ref{bc2d}).
After inserting (\ref{2Dto1D}) into (\ref{sg2d}) and integrating across
$y$ we get the evolution of $\phi_0$
\begin{equation}\label{phi0}
\phi_{tt} -\phi_{xx} + {w_j \over w} g(x,y=0)
(\kappa \phi_{tt} + \alpha \phi_{t} + \sin\phi  )= \nu {\gamma \over l}\;, 
\end{equation}
where we omitted the $0$ in $\phi_0$ and terms in $\phi_i, i \ge 1$
which are small because
of the smallness of the current \cite{bcf02}.

The boundary conditions are
\begin{equation}\label{bc1d}
{\phi}_x|_{x=0}= H-(1-\nu){\gamma \over 2},~~~~
{\phi}_x|_{x=l}= H+(1-\nu){\gamma \over 2},
\end{equation}
and
$$g(x) = \left\{
\begin{array}{l r}
\frac{w_j}{w}, & a_j-\frac{l_j}{2}<x<a_j+\frac{l_j}{2}, \\
0, & {\rm elsewhere}. \\
\end{array} \right..$$
The factor
$w_j/w$ is exactly the "rescaling" of $\lambda_J (=1)$ into
$\lambda_{eff} = \sqrt{w \over w_j}>1 $ due to the presence of the
lateral passive region \cite{cfv96}.

As the area of the junction is reduced the total super-current is
reduced and tends to zero. Small junctions where the
phase variation can be neglected and that have a significant
supercurrent we introduce the
following Dirac distribution model. First define the function
$g_h(x)$,
$$g_h(x) = \left\{
\begin{array}{l r}
\frac{w_j}{ w h}, & a_j-\frac{hl_j}{2}<x<a_j+\frac{hl_j}{2}, \\
0, & {\rm elsewhere}, \\
\end{array} \right.$$
notice that $g_1(x)=g(x)$.
When the junction widths $w_j \ll \lambda_J$ we can further reduce the problem
by neglecting the variation of the phase inside the junctions.
We then obtain the Dirac delta function distributed model \cite{cl05,louk_th}
by making $h$ tend to $0$. For an $n$ junctions device,
$$\lim_{h \to 0}g_h(x) = \sum_{j=1}^{n}\frac{w_j l_j}{l}\delta(x-a_j).$$
Finally we obtain the $\delta-1D$ model for the device,
\begin{equation}\label{delta-1D}
{\phi}_{tt} -{\phi}_{xx}+ \sum_{j=1}^{n} d_j \delta(x-a_j)
( \kappa {\phi}_{tt} + \sin {\phi} + \alpha {\phi}_{t})
=\nu \frac{\gamma}{l},
\end{equation}
where 
\begin{equation}\label{defdj}
d_j = \frac{w_j l_j}{w}
\end{equation}
and where the boundary conditions are given by eq.(\ref{bc1d}). 
This is the main model of the article and we will analyze it in
detail.

\section{Preliminary analysis}

From the Josephson equation (\ref{josephson}) it can be seen that
${\phi}_{t}$ is a voltage. In experiments this instantaneous voltage
is of very high frequency ($\sim$ 500 GHz) and can only be detected
by making it beat with a well-known source. On the other hand the time
average voltage can be measured in a fairly standard way. This
current voltage relation (the $I-V$ curve) is a characterization of the device. 
It is therefore important for the analysis to explain it. 
We now give some important symmetries of the $I-V$ curve. The first one
is the periodicity with respect to the magnetic field $H$. This is similar
to the one obtained in the static case \cite{cl06}.

\subsection{Periodicity of the $I-V$ curve with $H$}

The $I-V$ curve of the device modeled by eq.(\ref{delta-1D}) and 
(\ref{bc1d}) depends on the magnetic field $H$. We denote $I-V|_H$
the $I-V$ curve of the device for the magnetic field $H$.
Let us introduce $l_j=a_{j+1}-a_j$ the distance between two 
consecutive junctions. Let $l_{min}$ be the smallest distance 
$l_j$. We define a harmonic array as a circuit
where $l_i$ is a multiple of $l_{min}$ for all $i$. 

\paragraph{Proposition, Periodicity of the device:}
For a harmonic circuit, the $I-V|_H$ 
curve is periodic with a period $2\pi/l_{min}$.

\paragraph{Proof:}
Let $\phi$ be a solution of (\ref{delta-1D}) for a current $\gamma$ and a magnetic field $H$. We introduce $f(x)=(2\pi/l_{min})(x-a_1)$ and 
$\psi(x;t) = \phi(x;t) + f(x)$. So $\psi$ verifies
\begin{equation}
{\phi}_{tt} -{\phi}_{xx}+ \sum_{j=1}^{n}\frac{w_j l_j}{l}\delta(x-a_j)
( \kappa {\phi}_{tt} + \sin ({\psi}-f) + \alpha {\phi}_{t})
=\nu \frac{\gamma}{l},
\end{equation}
with $\psi_x(0;t) = H+2\pi/l_i - (1-\nu)\gamma/2$, and
$\psi_x(l;t) = H+2\pi/l_i + (1-\nu)\gamma/2$.  
Since, $f(a_j)=2k\pi$, $\forall i\in \{1;...;n\}$, then $\psi$ 
is a solution of (\ref{delta-1D}) for $H+H_p \equiv H+2\pi/l_{min}$ 
and for the same $\gamma$. 

Conversely, by subtracting $f$ from a solution associated to $H+H_p $
and a current $\gamma$, we obtain a solution for $H$ and the same 
current $\gamma$. We have shown that for a given $\gamma$ and $H$, 
if there is a solution, we can find another for $H+2k\pi/l_{min}$ 
($k \in {Z}$). We obtain the same $I-V$ curves. 
\begin{equation}\label{Prdct}
I-V|_{H+H_p} = I-V|_H.
\end{equation}
with $H_p = 2\pi/l_{min}$.
\qquad

In the non harmonic case, if the junctions are set such that
$l_j=p_j/q_j$, where $p_j$ and $q_j$ are integers, prime with 
each other, then  $I-V|_H$ is periodic with period $H_p$ such that
\begin{equation}\label{periodicity}
H_p = 2\pi \frac{LCM(q_1;...;q_{n-1})}{HCF(p_1;...;p_{n-1})},
\end{equation} 
where $LCM$ is the Lowest Common Multiple and $HCF$ the 
Highest Common Factor. To prove this write 
$f(x) = H_p (x-a_1)$ and use again the previous argument.

\subsection{The high voltage approximation }

When the voltage $\phi_t$ is large, the phase $\phi$ is
rotating fast so that one can write 
\begin{equation}\label{phi_hv} \phi(x,t) =
V t + \psi(x,t),\end{equation}
where the average 
\begin{equation}\label{phiv} \left<\psi \right>
\equiv {1\over T}\int_t^{t+T} \psi(x,t') dt'
\equiv \phi_v(x).\end{equation}
Plugging the ansatz (\ref{phi_hv}) into (\ref{delta-1D}) and taking
the average we get
\begin{eqnarray}
\left<\psi_{tt} \right>-{\phi_v}_{xx} + \sum_{j=1}^n  d_j \delta(x-a_j)
&&\nonumber \\
[-1.5ex] \label{av_delta-1D} \\ [-1.5ex]
\left[  \kappa \left<\psi_{tt} \right> 
+ \left< \sin(V t) \cos(\psi) \right>
+\left< \cos(V t) \sin(\psi)\right> + 
\alpha \left< \psi_{t} \right> + \alpha V \right] &=&\gamma. 
\nonumber
\end{eqnarray}
Then if we neglect the nonlinear terms we obtain 
the static equation, we obtain a new equation such 
that $\left<\psi_{tt} \right> =\left<\psi_{t} \right>=0$ is a
solution. Thus,
\begin{equation}\label{hvolt}
 -{\phi_v}_{xx} + \alpha V \sum_{j=1}^n  d_j \delta(x-a_j)
= \nu {\gamma \over l} , \end{equation}
together with the boundary conditions
\begin{equation}\label{bchvolt}
{\phi_v}_x|_{x=0} = H - (1-\nu) {\gamma \over 2},~~
{\phi_v}_x|_{x=l} = H + (1-\nu) {\gamma \over 2} .\end{equation}
By integrating the equation (\ref{hvolt}) one sees that
this problem has a solution if
\begin{equation}\label{iv}
V = {\gamma \over  \alpha \sum_j d_j}.
\end{equation}

Let us write this high voltage solution for a device with many junctions.
At each junction $x=a_j$, the phase must be continuous. Integrating equation 
(\ref{hvolt}) over a neighborhood of $a_j$ and taking the limit of this neighborhood
going to zero one gets the jump condition
\begin{equation}\label{jump_phiv} 
-\left[{\phi_v}_x\right]_{a_j^-}^{a_j^+} 
+ \alpha V {d_j \over l}=0.
\end{equation}
Notice also that outside the junctions, the solution has to be of the form
$$\phi_v(x) = -\nu {\gamma \over 2l}x^2 + c_1 x + c_2.$$
It is then natural to build the solution by steps, using a sort of
shooting method. To simplify the discussion, let us consider a device with two
junctions placed respectively at the positions $a_1$, $a_2$. We write
\begin{equation}\label{shoot_phiv}
\phi_v(x) = \left\{ \begin{array}{ll}
\phi_0(x),&~0\le x\le a_1,\\
\phi_1(x),&~a_1\le x\le a_2,\\
\phi_2(x),&~a_2\le x\le l.
\end{array} \right. 
\end{equation}
At each junction $a_j$ we have $\phi_{j+1}(a_j)=\phi_j(a_j)$ and
$${\phi_{j+1}}_x(a_j^+)- {\phi_j}_x(a_j^-) +\alpha V {d_j \over l}=0.$$
It is then natural to write 
\begin{equation}\label{phiv0}
\phi_0(x) = -\nu {\gamma \over 2l}x^2 + \left[H-(1-\nu)\frac{\gamma}{2}
\right]x + C.
\end{equation}
From the relations at each junction, we infer that 
\begin{eqnarray}
\phi_1(x) = \phi_0(x) + \gamma {d_1 \over d_1+d_2}(x-a_1),
\nonumber \\
[-1.5ex] \label{phiv12} \\ [-1.5ex]
\phi_2(x) = \phi_1(x) + \gamma {d_2 \over d_1 + d_2}(x-a_2).
\nonumber 
\end{eqnarray}
It can be checked that the right boundary condition for $x=l$ can verified
by this formulation whatever the value of $H$ by choosing $\gamma$ thanks to
eq.(\ref{iv}). 
The function $\phi_v(x)$ defined by the equations (\ref{phiv0}-\ref{phiv12})
is then the solution of the problem (\ref{hvolt},\ref{bchvolt}).

\subsection{Simplification of the $\delta-1D$ model}

The static part $\phi_v$ of the high voltage solution can be used
to simplify the formulation of the problem, in particular the boundary
conditions. For that we introduce
\begin{equation}\label{psi} 
\psi(x,t) = \phi(x,t) -\phi_v(x),
\end{equation}
so that the sine-Gordon equation (\ref{delta-1D}) becomes
\begin{equation}\label{delpsi_phiv}
\psi_{tt} -\psi_{xx} + \sum_{j=1}^n  d_j \delta(x-a_j)
\left [ \kappa \psi_{tt}+ \alpha \psi_{t} + \sin(\psi + \phi_v) 
- \frac{\gamma}{\sum_{k=1}^n d_k}
\right ]= 0\;,\end{equation}
with the homogeneous Neumann boundary conditions
\begin{equation}\label{bcpsi}
\psi_x|_{x=0} = 0,~~\psi_x|_{x=l} = 0 . \end{equation}
Equation (\ref{delpsi_phiv}) now contains for each junction 
a sine term with an argument
that is shifted by $\phi_v(a_j)-\phi_v(a_1)$. When averaging
over time we get
$$ \left< \phi(a_j) -\phi(a_1) \right>_t  
\approx \phi_v(a_j) - \phi_v(a_1)\;.$$
We define $\Psi_j$ as the phase-shift of the junction $j$
\begin{equation}\label{def_dephas}
\Psi_j=\phi_v(a_j) - \phi_v(a_1)\;.
\end{equation}
By a simple translation, equation (\ref{delpsi_phiv}) becomes
\begin{equation}\label{delpsi}
\psi_{tt} -\psi_{xx} + \sum_{j=1}^n  d_j \delta(x-a_j)
\left [\kappa \psi_{tt} + \alpha \psi_{t} + \sin(\psi + \Psi_j) -
\frac{\gamma}{\sum_{k=1}^n d_k}\right ]= 0\;.\end{equation}
The boundary conditions (\ref{bcpsi}) are unchanged and $\Psi_1=0$.

\section{Resonances for an array of point junctions}

\subsection{Spectral problem}

When the system is on a resonance, the solution is periodic so that
it can be written as $\psi(x,t) = e^{i \omega t} \varphi(x)$. In that
case the terms $\alpha \psi_{t} + \gamma/\sum_{k=1}^n d_k$
globally balance each other. The sine term can be averaged out. Therefore
to satisfy the equation (\ref{delpsi}) it is necessary that
\begin{equation}\label{linpsi}
\psi_{tt} -\psi_{xx} + \sum_{j=1}^n  d_j \delta(x-a_j)
\kappa \psi_{tt}  = 0\;,\end{equation}
together with the homogeneous Neumann boundary conditions (\ref{bcpsi}).

When the capacities per unit area of the junctions and passive regions
are equal $\kappa=0$, so the array resonates on a cosine Fourier mode,
see \cite{cl05,cl07}.
$$\psi_n(x,t) = e^{i  {n\pi t \over l}}
\cos\left( {n\pi x \over l}\right).$$
In this description, we have neglected the higher harmonics which 
decay exponentially as shown by numerical calculations.
When $\kappa\neq 0$, the eigenmodes of equation (\ref{linpsi}) 
differ from $n \pi /l$. To analyze them we substitute 
$\psi(x,t) = e^{i \omega t} \varphi(x)$ into (\ref{linpsi}) and obtain
the following eigenvalue problem
\begin{equation}\label{eigvp}
\varphi_{xx} + \omega^2 \left [ 1+ 
\sum_{j=1}^n  d_j \delta(x-a_j) \kappa \right] \varphi =0,\end{equation}
together with homogeneous Neumann boundary conditions for $\varphi$.

We consider the two junctions case.
We will obtain results that can be generalized to an array with
$n>2$ junctions. We introduce for ease of notation 
$\kappa_j = \kappa d_j$.
It is then natural to assume a solution
\begin{equation} 
\varphi(x) = \left\{ \begin{array}{lll}
A_1 \cos(\omega x),&{\rm for}& 0\le x \le a_1,\\
A_2 \cos(\omega x) + B_2 \sin(\omega x), 
& {\rm for} & a_1\le x \le a_2,\\
A_3 \cos(\omega x) + B_3 \sin(\omega x), & {\rm for} & a_2\le x \le l,
\end{array} \right .
\end{equation}
where the form of $\varphi$ in the first interval was chosen to satisfy the
boundary condition at $x=0$.
As usual $\varphi$ is continuous at the junctions
$x=a_j$ and one can see that the jump of the derivative is 
\begin{equation}\label{jvpx} 
[\varphi_x]_{a_j^-}^{a_j^+}+\kappa_j \omega^2 \varphi(a_j)=0.
\end{equation}
Writing these $2 n = 4 $ conditions at the junctions and the 
boundary condition
at $x=l$, one obtains a 5th order homogeneous system in $A_1,A_2,B_2,
A_3,B_3$. The solution is non trivial if the following determinant is zero
\begin{equation}\label{disp}
\left \| \begin{array}{lcccr}
C_1   &  -C_1  & -S_1  & 0 & 0   \\
S_1   &  - S_1 +\kappa_1 \omega C_1 & C_1 + \kappa_1 \omega S_1  & 0 & 0   \\
0   & C_2  &   S_2      & -C_2  & -S_2    \\
0 & S_2 & -C_2 & -S_2 +\kappa_2 \omega C_2   & C_2+\kappa_2 \omega S_2    \\
0     & 0       &   0      & -S_l   & C_l    
\end{array} \right \|  =0, 
\end{equation}
where 
$C_1=\cos(\omega a_1),~S_1=\sin (\omega a_1),
~C_2=\cos(\omega a_2),~S_2=\sin (\omega a_2),
~C_l = \cos(\omega l),~S_l = \sin (\omega l).$
When  more junctions are present in the device, the determinant giving the
dispersion relation can be generated by adding the elementary component
in rows 3 and 4 corresponding to each additional junction.

We now give the resonant frequencies when there is only one junction.
The determinant (\ref{disp}) becomes
\begin{equation}\label{detdisp1}
\left \| \begin{array}{lcr} C_1   &  -C_1  & -S_1     \\
S_1   &  - S_1 +\kappa_1 \omega C_1 & C_1 + \kappa_1 \omega S_1 \\
0     & -S_l   & C_l
\end{array} \right \|  =0,
\end{equation}
which implies 
\begin{equation}\label{dispersion}
\sin(\omega l) + \kappa \omega d_1 \cos(\omega a_1) \cos(\omega (l-a_1)) =0.\end{equation}
Several remarks should be made on this relation. \begin{itemize}
\item First, when
$\kappa \ll 1$ we recover the usual $\sin(\omega l)=0$ dispersion
leading to harmonic frequencies. 
\item The opposite limit $\kappa \gg 1$
is more interesting because it leads to a splitting of the oscillations
in the left and in the right side of the film. We obtain 
$\cos(\omega a_1)=0$ or $\cos(\omega (l-a_1)) =0$ leading to
$\omega a_1 = (2n+1) \pi/2$ or $\omega (l-a_1)= (2m +1)\pi/2$ where
$m,n$ are integers.
\end{itemize}
\begin{figure}
\centerline{\epsfig{file=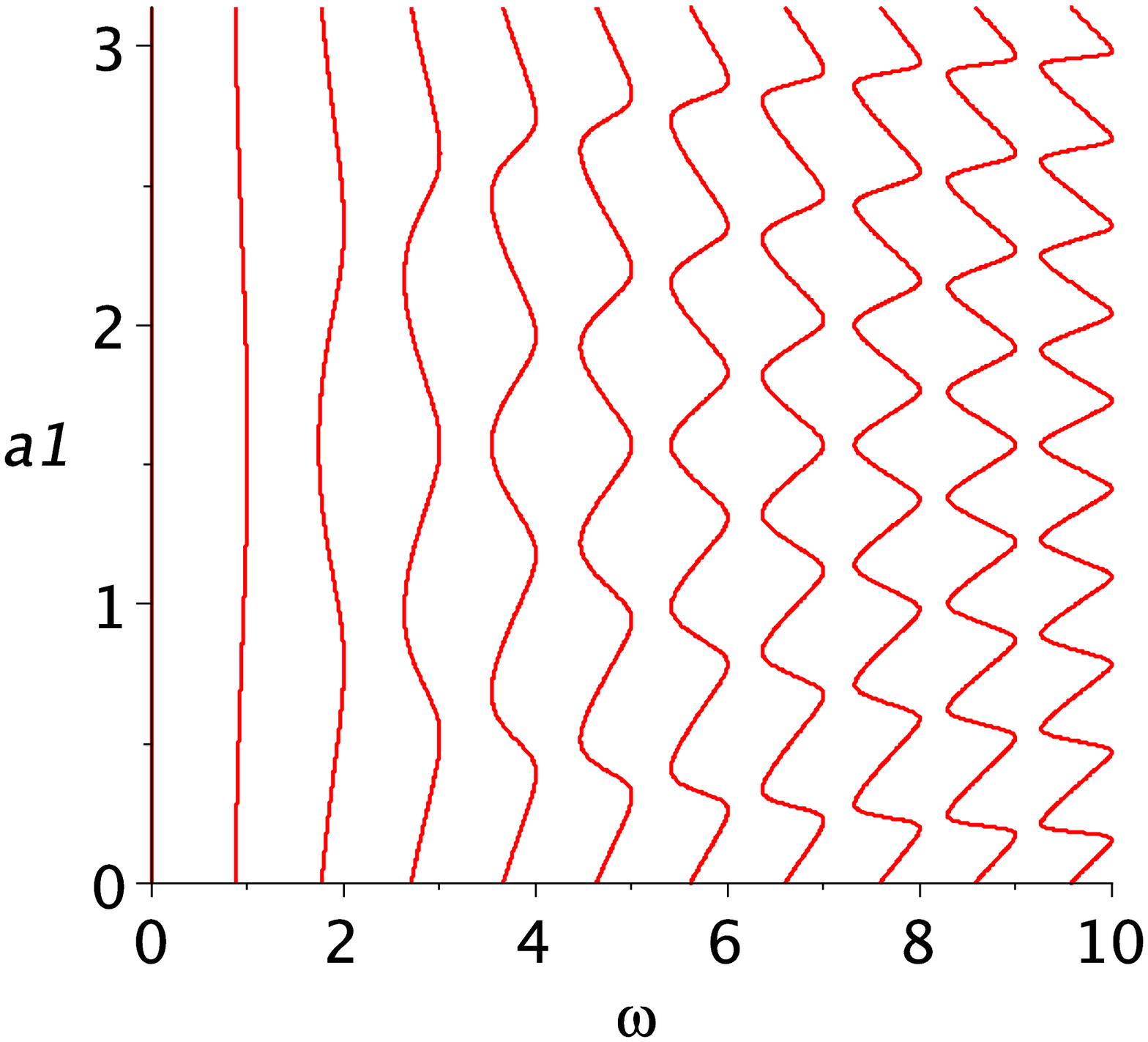,width=0.5\linewidth,angle=0}
\epsfig{file=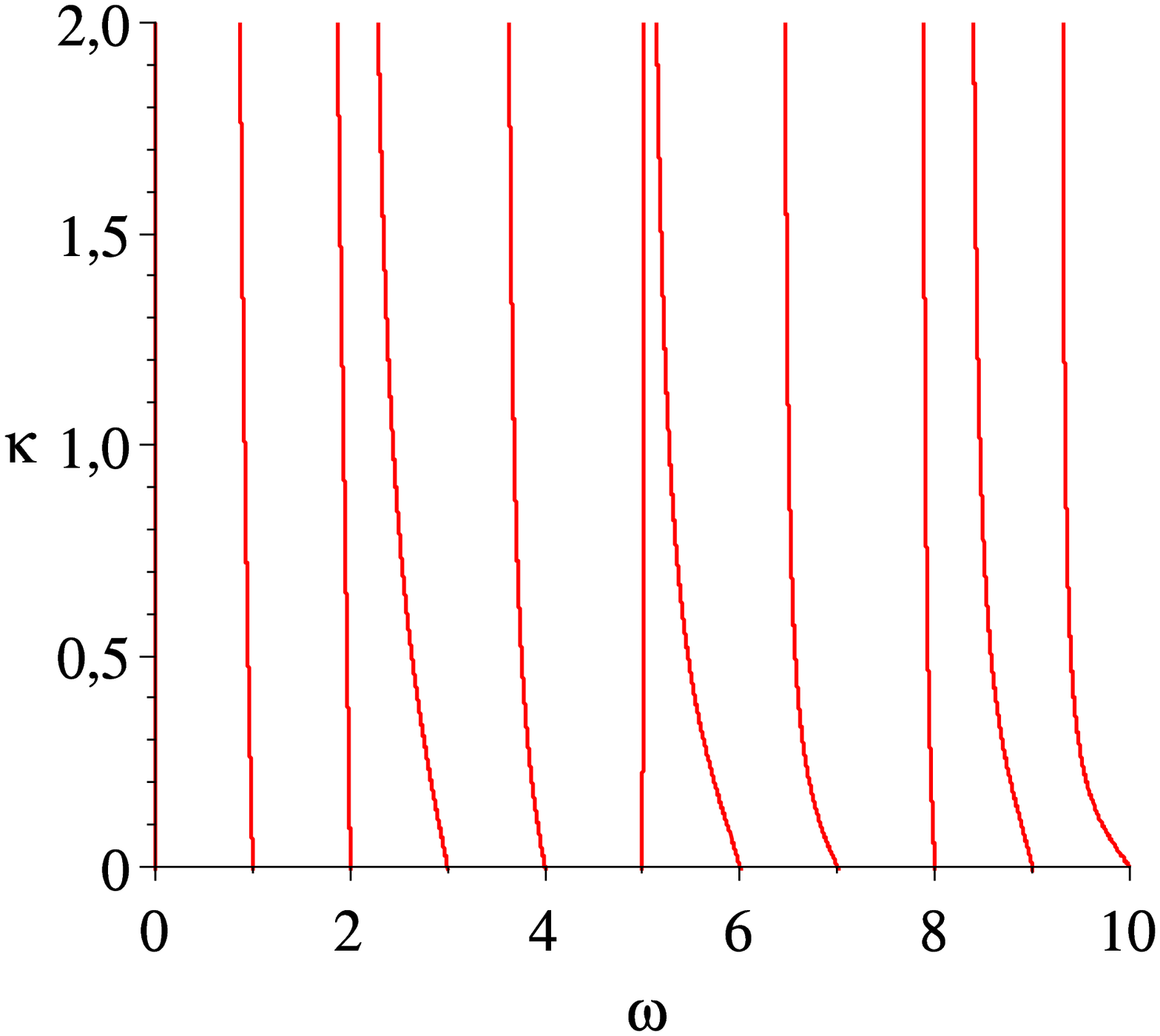,width=0.5\linewidth,angle=0}}
\caption{Dependence of the eigenvalues (zeros of the dispersion relation
(\ref{dispersion}) ) on the position of the junction
$a_1$ in the microstrip (left panel) and on the capacity miss-match $\kappa$
(right panel). For the left panel, $\kappa=0.5$. For the right panel 
$a_1=0.3 l = 0.3 \pi$.}
\label{f2}
\end{figure}

Notice that Larsen et al \cite{ldm91} considered
a centered junction $a_1=l/2$ in a microstrip. Their dispersion relation 
$$ \tan \pi  {f \over f_e} = - \pi {C_j \over C_e} {f \over f_e},$$
is similar to the one we get with our approach except
that the coefficient is different. We obtain using (\ref{dispersion})
\begin{equation}\label{disp_larsen}
\tan (\omega l) = - \omega l  \left({ C_j \over C_l} -1\right) 
{w_j^2 \over w l}.\end{equation}
This latter expression gives
the correct eigenmodes as shown in the current voltage characteristics
computed numerically shown in the next sections.

\subsection{Eigenvectors and inner product}

We introduce here the inner product associated with the dispersion
relation (\ref{dispersion}).
The eigenvectors $\varphi_i$ (respectively
$\varphi_j$) associated to the eigenvalue $\omega_i$ (resp.
$\omega_j$) satisfy
\begin{eqnarray}\label{varepsi_i}
{\varphi_i}_{xx}+\omega_i^2 \varphi_i
\left(1+\delta(x-a_1)\kappa_1\right)&=&0,\\
\label{varepsi_j}
{\varphi_j}_{xx}+ \omega_j^2 \varphi_j
\left(1+\delta(x-a_1)\kappa_1\right)&=&0,
\end{eqnarray}
together with the boundary conditions 
\begin{equation}\label{bc_varpsi}
\left. {\varphi_i}_{x}\right|_{x\in \{0,l\}}=0,
~{\rm and}~~\left. {\varphi_j}_{x}\right|_{x\in \{0,l\}}=0.
\end{equation}
We assume the eigenvalues to be different.
As usual we multiply eq.(\ref{varepsi_i}) by $\varphi_j$ and
eq.(\ref{varepsi_j}) by $\varphi_i$, substract the second
from the first and integrate on the domain. We obtain
\begin{equation}
\int_0^l ({\varphi_i}_{xx} \varphi_j-
{\varphi_j}_{xx} \varphi_i) dx
+(\omega_i^2-\omega_j^2)
\left [
\int_0^l\varphi_i\varphi_jdx
+ \kappa_1\varphi_i(a_1) \varphi_j(a_1)  
\right ] =0.\end{equation}
The first integral is zero because of the boundary conditions 
(\ref{bc_varpsi}). Since $\omega_i\neq \omega_j$ we get 
\begin{equation} \int_0^l (1+ \kappa_1\delta(x-a_1))
\varphi_i\varphi_j dx =0. \label{subs_lin}
\end{equation}
Thus we obtain the orthogonality of the eigenvectors $\varphi_i$,
$i$ is an integer with the associated inner product defined by
\begin{equation}\label{scl}
\left< f;g\right> \equiv  \int_0^l \left(1+ \kappa_1\delta(x-a_1)
\right)fg dx.
\end{equation}
It is then possible to normalize the eigenvectors so that they form
an orthonormal basis. When there are $n$ junctions in the array, the
inner product can be generalized easily to 
\begin{equation}\label{scln}
\left< f;g\right> \equiv  
\int_0^l \left(1+ \sum_{i=1}^n\kappa_i\delta(x-a_i)\right)fg dx.
\end{equation}

\begin{figure}
\centerline{
\epsfig{file=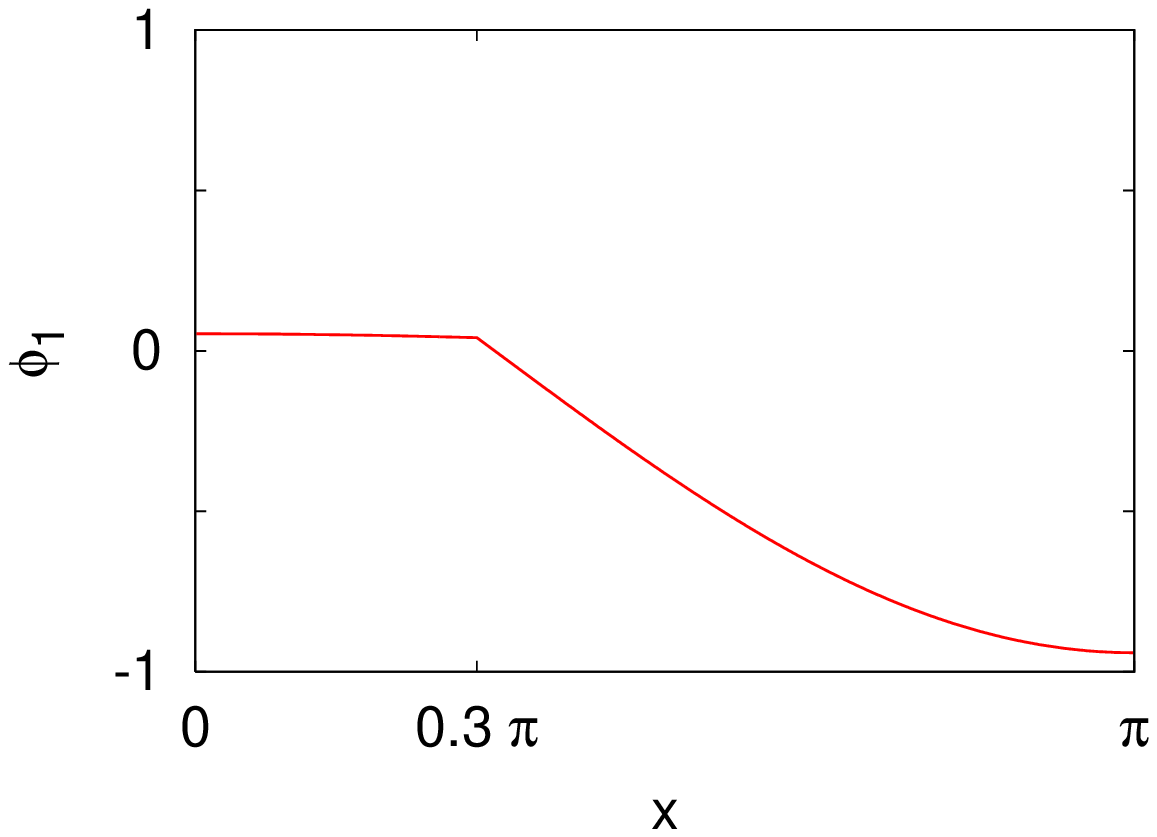,width=0.5\linewidth,angle=0}
\epsfig{file=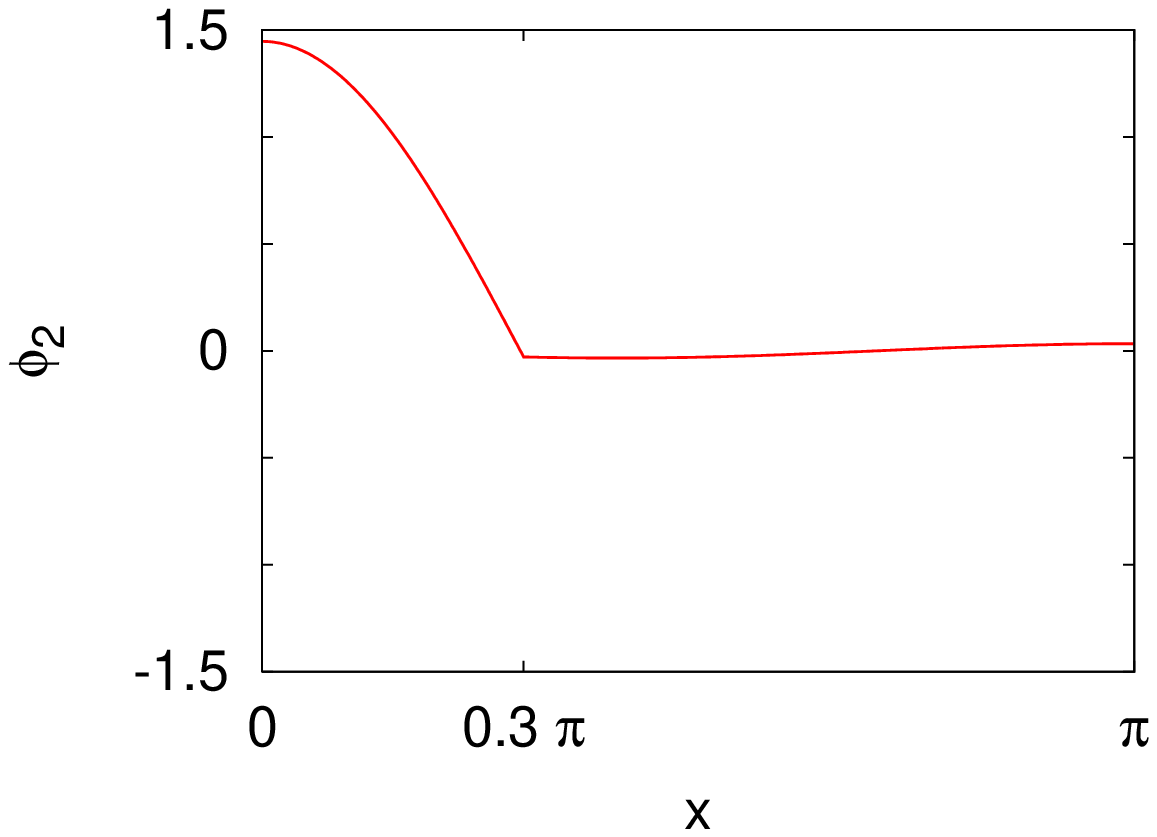,width=0.5\linewidth,angle=0}}
\centerline{
\epsfig{file=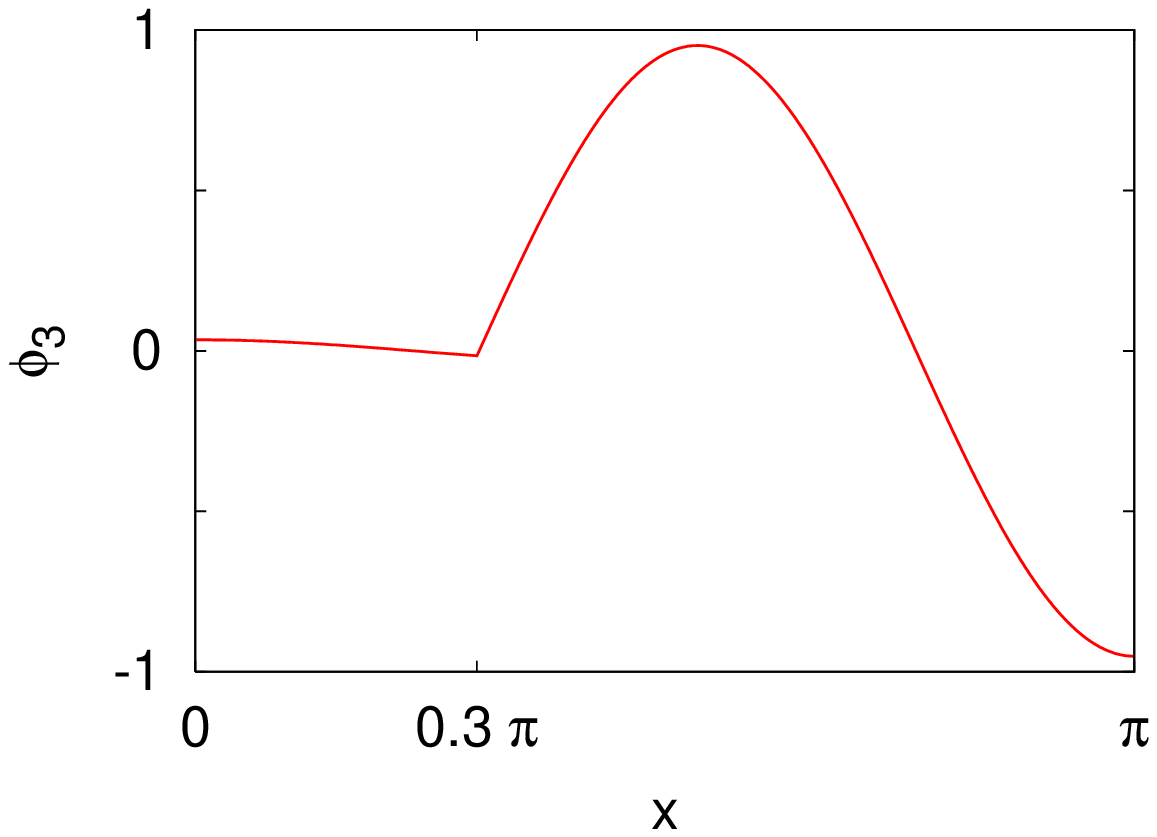,width=0.5\linewidth,angle=0}
\epsfig{file=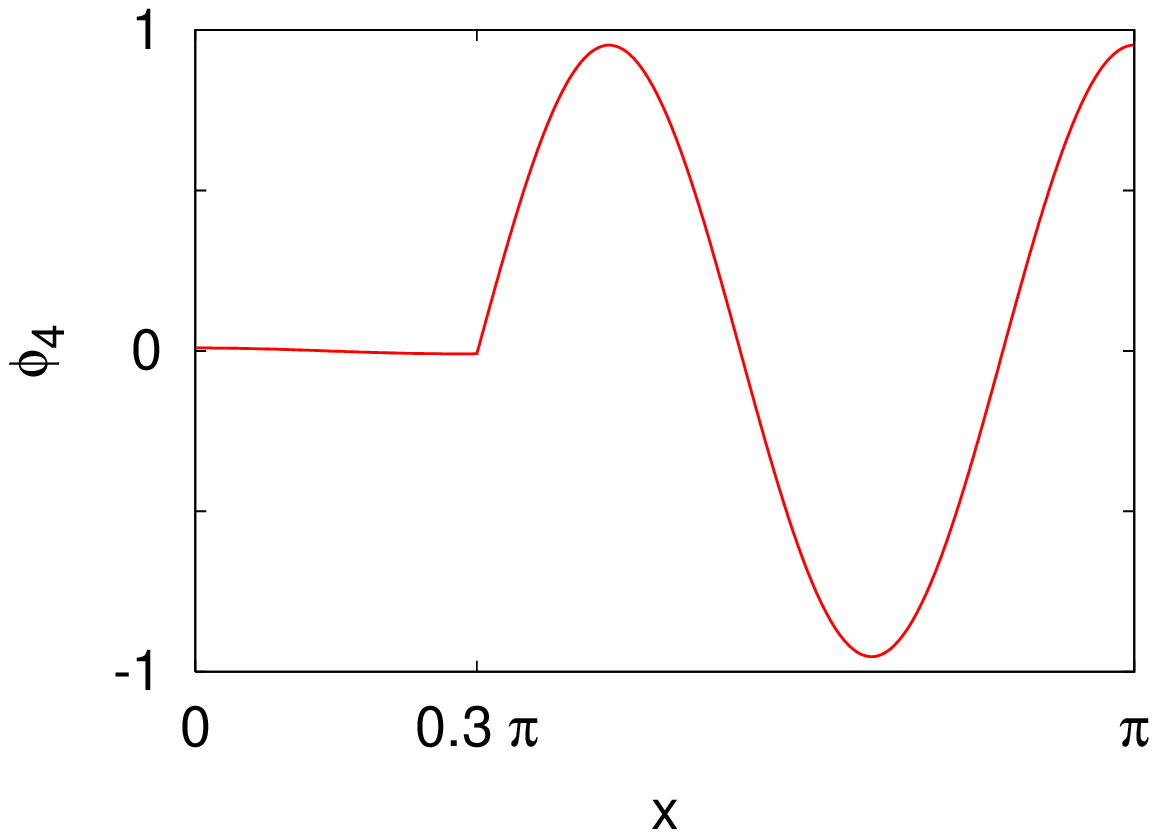,width=0.5\linewidth,angle=0}}
\caption{Normalized eigenmodes $\varphi_i(x), i=1,2 $ 
(top panels from left to right)
and $\phi_i(x), i=3,4 $ (top panels from left to right) for a
large capacity miss-match $\kappa=30$. The other parameters are
$d_1=1$, $l=\pi$ and 
$a_1=0.3\pi$.}
\label{f3a}
\end{figure}

In the case of a single junction, 
the normalized eigenvectors $\varphi_n(x) $ are given by
\begin{equation} \label{phin}
\varphi_n(x) = \left\{\begin{array}{lll}
A_n \cos(\omega_n x),& {\rm for}& 0\le x \le a_1,\\
A_n {\cos(\omega_n a_1) \over\cos(\omega_n(l- a_1))} 
\cos(\omega_n(l- x)), & {\rm for} & a_1\le x \le l,
\end{array} \right.
\end{equation}
where $\omega_n$ satisfies the dispersion relation (\ref{dispersion})
and 
\begin{equation}\label{an}
A_n = \frac{1}{\sqrt{{l \over 2} +{\sin 2 \omega_n a_1 \over 4 \omega_n }
+{\cos^2 \omega_n a_1 \over \cos^2 \omega_n (l-a_1)} 
{\sin 2 \omega_n (l-a_1) \over 4 \omega_n } 
+\kappa_1 \cos^2 \omega_n a_1 }}.
\end{equation}
Fig. \ref{f3a} shows the 4 non trivial eigenmodes $\phi_i,~i=1-4$ for 
a large capacity miss-match. Notice how the modes are almost zero
on one side of the cavity.

\subsection{Eigenvalues and $I-V$ curves}

The eigenvalues calculated in the previous section appear as
resonances in the current-voltage characteristics of the device.
We computed the $I-V$ curves from the numerical solution of
equation (\ref{delta-1D}). The singular partial differential
equation is integrated over reference volumes and the time
advance is done by an ordinary differential equation solver
(see Appendix \ref{appendix_num} for more details). The average voltage
$V\equiv \left<\phi_t\right>$ is computed over a time interval
$10/\alpha$ after waiting a time $100/\alpha$ for the solution
to stabilize. When the system is locked on a resonance, it oscillates
periodically on an eigenfrequency solution of the dispersion relation
eq.(\ref{dispersion}). We will show that the numerical solution
follows closely the corresponding eigenvector. 
\begin{figure}
\centerline{ \epsfig{file=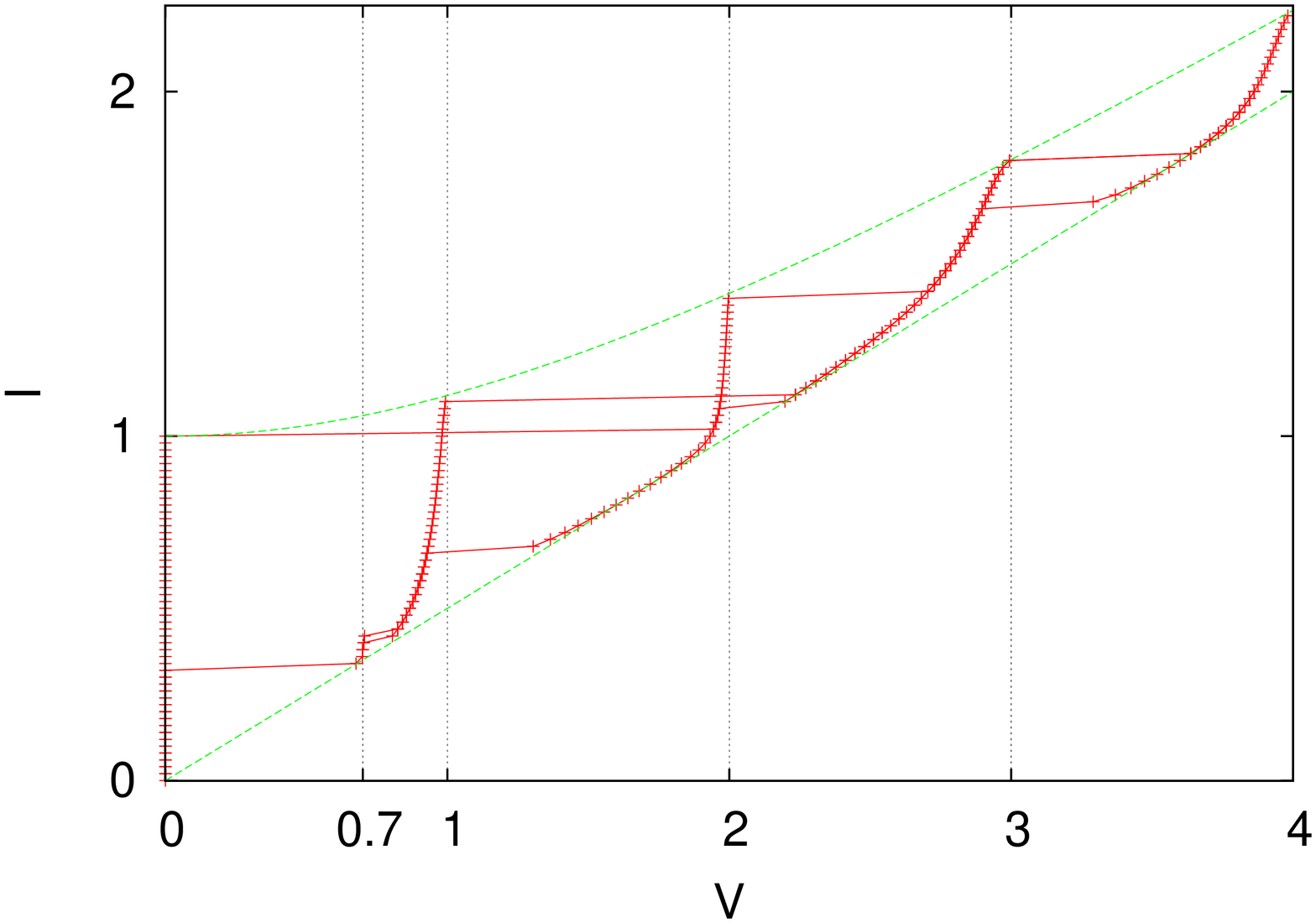,height=0.5\linewidth,angle=270}
\epsfig{file=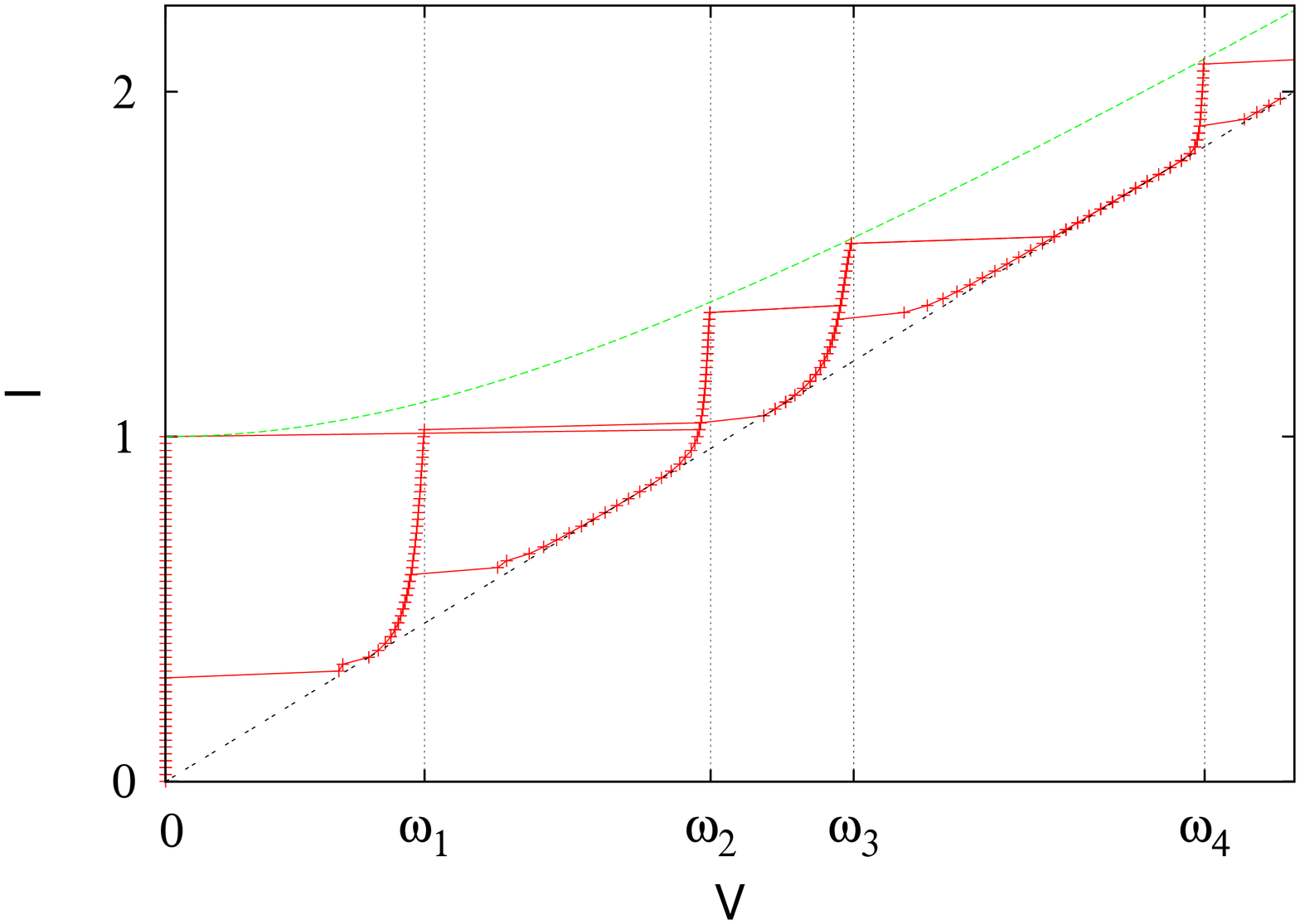,height=0.5\linewidth,angle=270}}
\centerline{ \epsfig{file=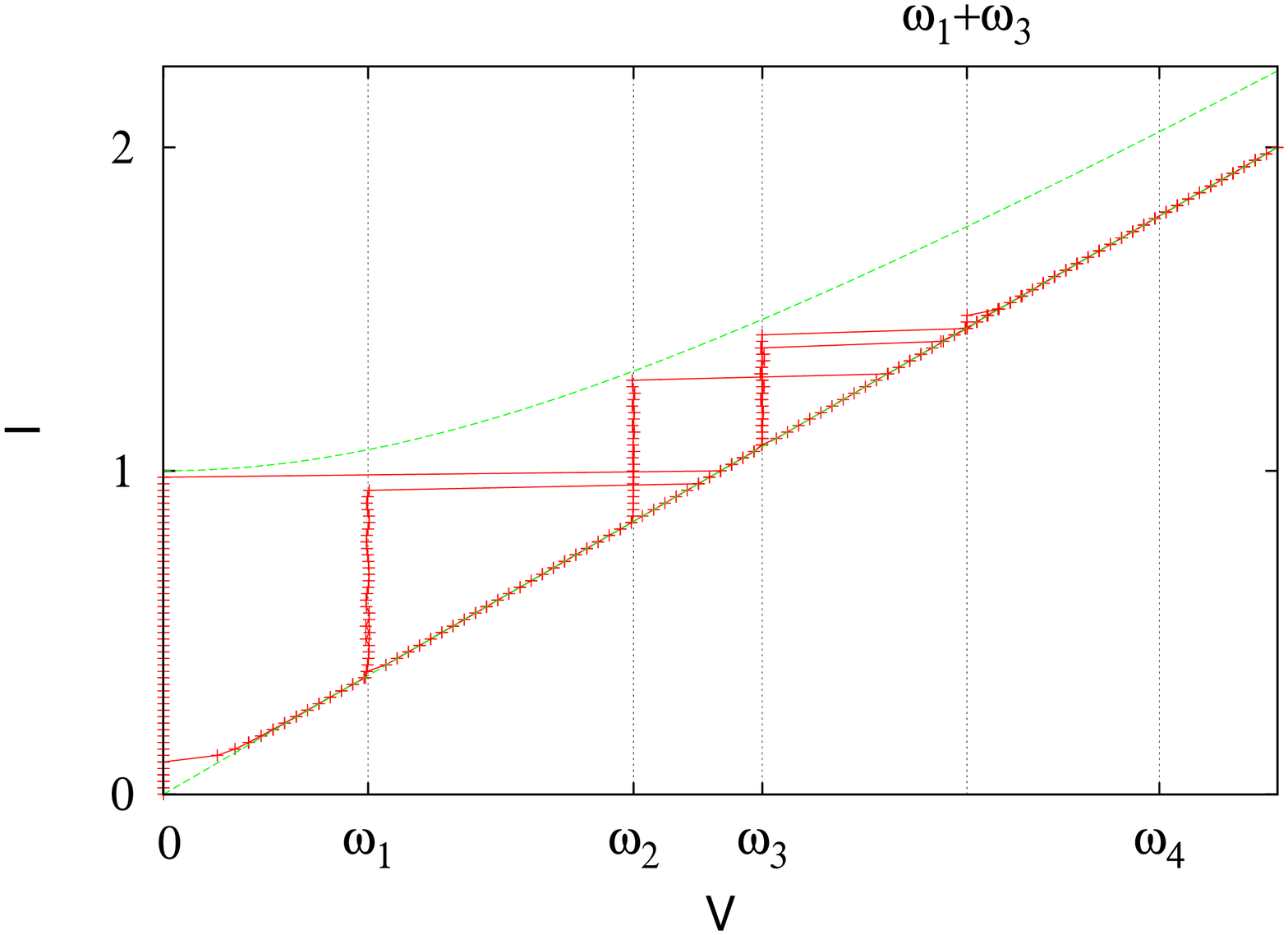,height=0.5\linewidth,angle=270}}
\caption{$I-V$ curves for an array with one junction, 
$d_1=1$, $\alpha=0.5$, $l=\pi$,
$a_1=0.3\pi$. Top panel, $\kappa=0$, middle panel $\kappa=1$ and
bottom panel $\kappa=30$.}
\label{f4}
\end{figure}
We choose a device with $l=\pi,~ a_1=l/3,~d_1=1$ and take three limiting
cases $\kappa=0$  $\kappa=1$ and $\kappa=30$. When there is no capacity
miss-match ($\kappa=0$) we showed in
\cite{cl05} that the resonances of the $I-V$ curve are positioned at 
$V=k\pi/l$ with $k$ integer and are
bounded by
$$\gamma l= d_1 \alpha V,~{\rm and}~~\gamma l=d_1 \sqrt{(\alpha V)^2+1}.$$
This is shown in the top panel of Fig. \ref{f4}. In the middle
panel we show the $I-V$ curve for $\kappa=1$. Notice how the resonances
are not equally spaced and get sharper. The resonances are very close
to the ones given by the dispersion relation, they are reported in
table (\ref{disp_1j_k1}). Finally on the bottom panel
we computed the $I-V$ curve for $\kappa=30$. This large value is
typical in experiments where the oxide layer in the junction is
about 100 times thinner than in the passive region. This together
with the ratio $d_1 = w_j^2/w$ gives $\kappa_1 \approx 50$.
Notice how the
resonances are vertical indicating that the system is almost
linear. This is to be expected because $\kappa \gg 1$ so that the
sine term plays little role.
The position of the resonances is exactly given by the
dispersion relation (\ref{dispersion}). They are reported in the table
(\ref{disp_1j}) together with the coupling coefficient
$\phi_n(a_1)$. The eigenvalues are clearly not harmonic.
Notice how some
resonances do not go all the way to the junction curve. Another observation
is that
we have a resonance for $\omega_2$ even though the coupling coefficient
is almost zero. We will come back to this point in the next section.
Also it is interesting that there is no resonance for $\omega_4$.
\begin{table}\label{disp_1j_k1}
$$\begin{array}{|c|c|c|c|} \hline
i & \omega_i & \cos(\omega_i a) & {\rm Normalization} ~ A_i\\ \hline
0 &  0.      &       1.           &     0.491    \\ \hline
1 &  0.917   &  0.649             &     0.562  \\\hline    
2 &  1.931   & -0.247             &    1.094    \\ \hline
3 & 2.438    & -0.665              &    0.704   \\ \hline     
4 & 3.682    & -0.946              &    0.237   \\ \hline     
\end{array}$$
\caption{Eigenvalues for the array of Fig. \ref{f4} with $\kappa=1$
(middle panel of Fig. \ref{f4}).}
\end{table}
\begin{table}\label{disp_1j}
$$\begin{array}{|c|c|c|c|} \hline
i & \omega_i & \cos(\omega_i a) & {\rm Normalization} ~ A_i\\ \hline
0 &  0.      &       1.           &     0.1737    \\ \hline
1 &  0.734   &  0.770             &     5.341 ~10^{-2}\\\hline
2 &  1.687   & -1.950~10^{-2}      &    1.447       \\ \hline
3 & 2.150    & -0.440              &    3.462~10^{-2}\\ \hline
4 & 3.575    & -0.974              &    9.102~10^{-3}\\ \hline
\end{array}$$
\caption{Eigenvalues for the array of Fig. \ref{f4} with $\kappa=30$
(bottom panel of Fig. \ref{f4}).}
\end{table}

\begin{figure}
\centerline{ \epsfig{file=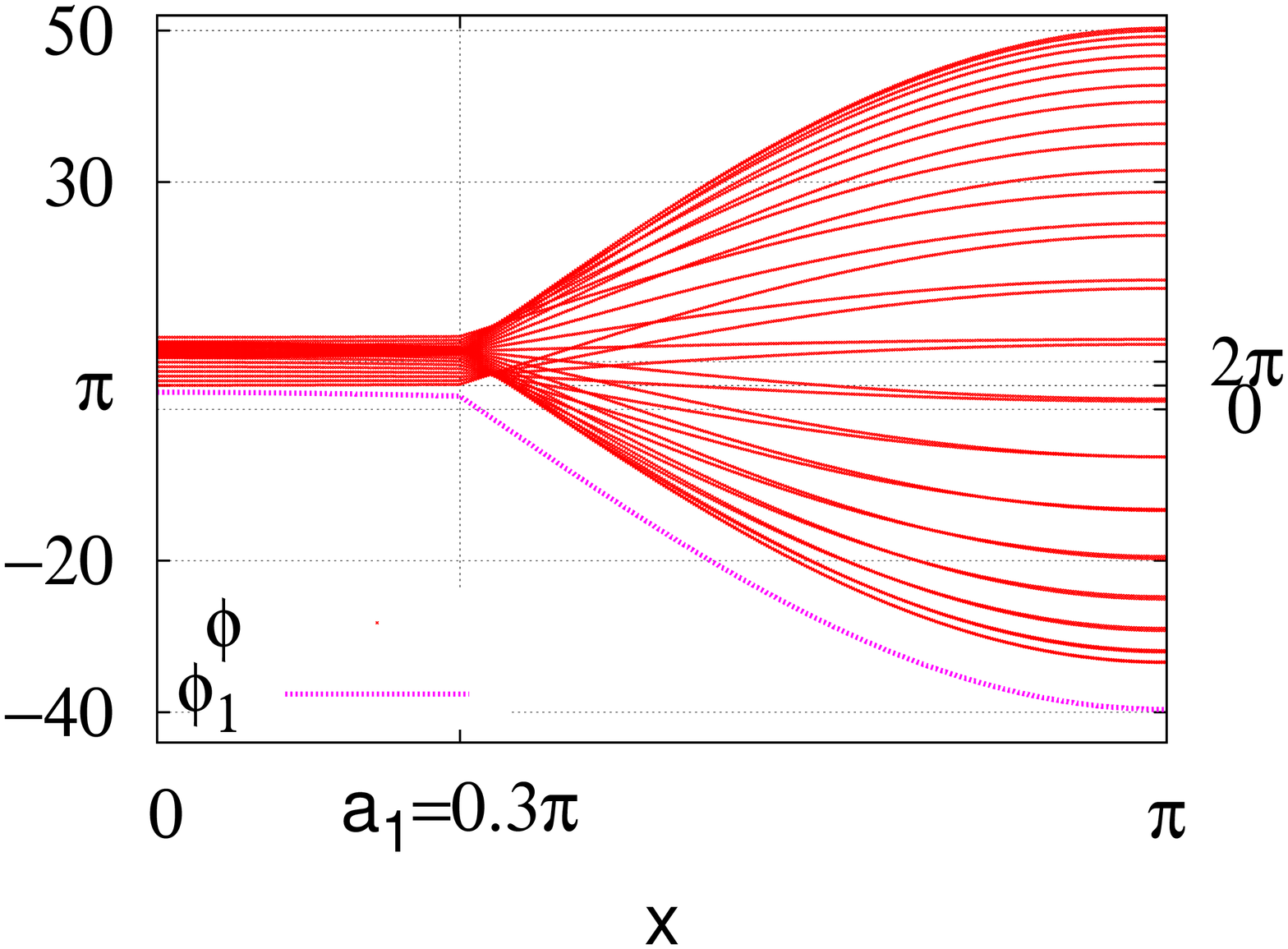,height=0.5\linewidth,angle= 270}
\epsfig{file=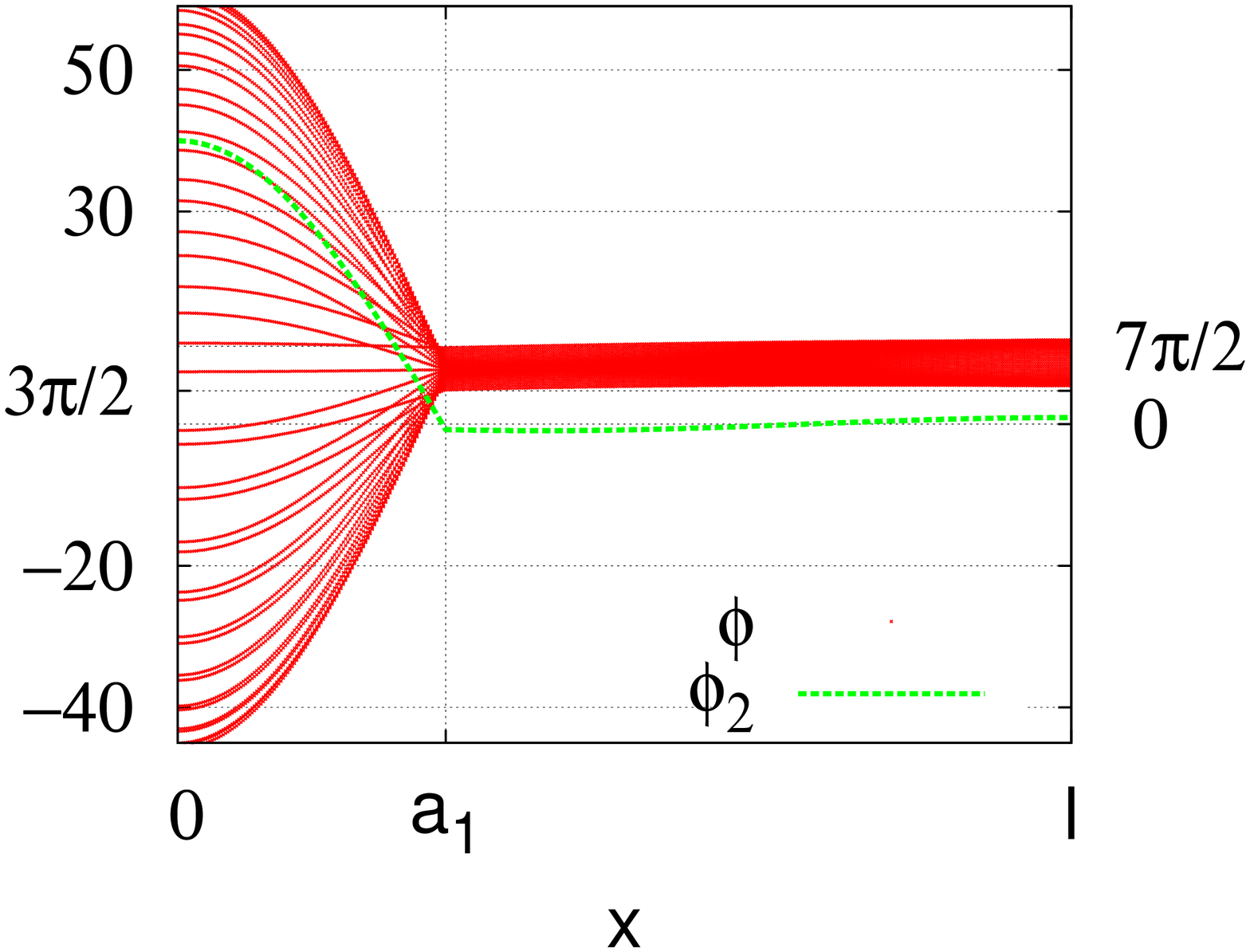,height=12 cm,height=0.5\linewidth,angle= 270}}
\centerline{\epsfig{file=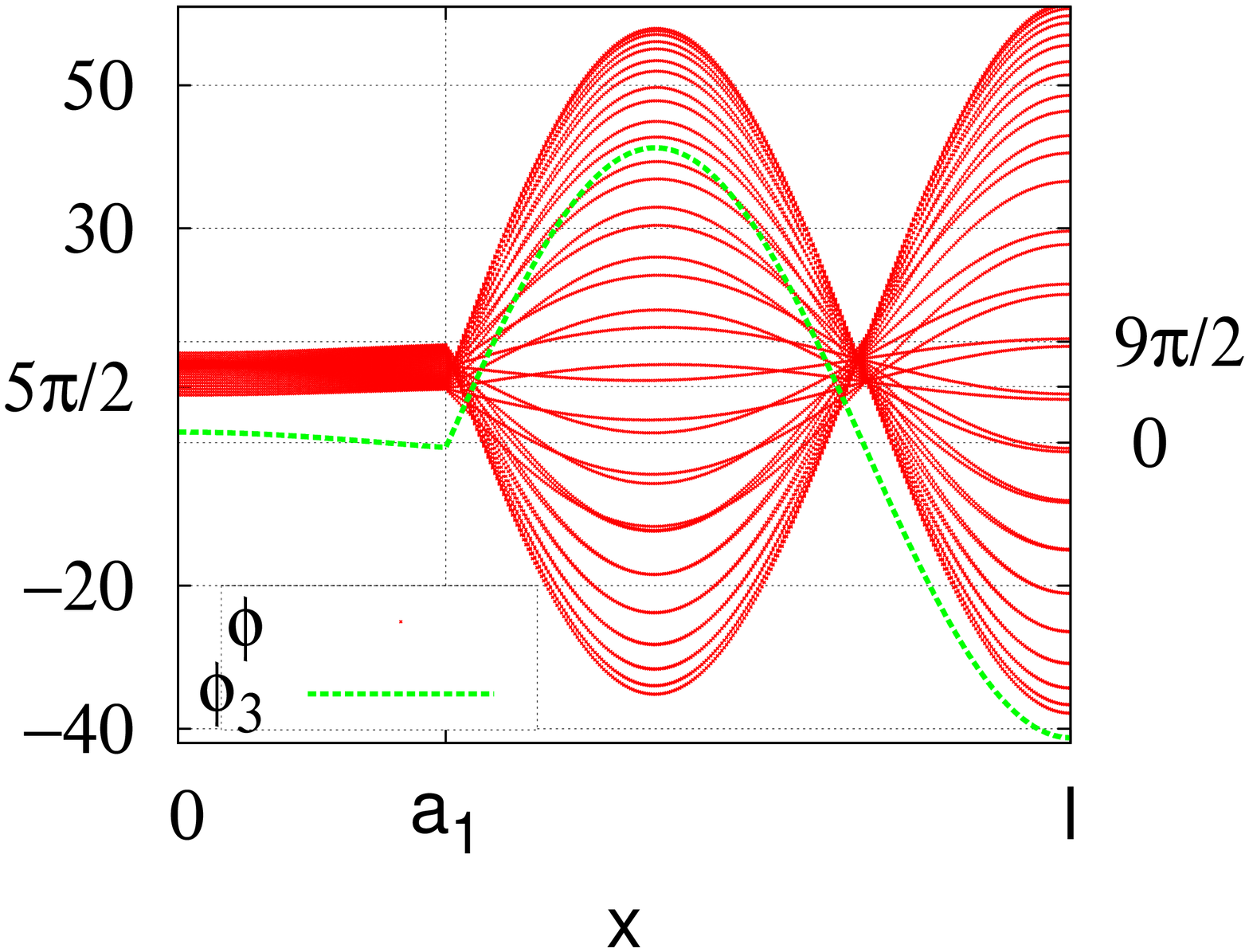,height=0.5\linewidth,angle= 270} }
\caption{Plot of the instantaneous voltage $\phi_t(x)$ for successive times
for the first, second and third resonances in the $I-V$ curve 
for an array with one junction and with a large 
capacity miss-match $\kappa_1=30$. The corresponding eigenmodes
(\ref{phin}) are plotted in dashed lines.}
\label{f4a}
\end{figure}

Fig. \ref{f4a} shows the eigenvectors in dashed line (green online) for
a realistic value $\kappa_1=30$ corresponding to a capacity
miss-match $\kappa=30$ and a surface ratio $d_1=w_1~l_1 /w= 1$.
On the same plot are indicated in continuous line (red online) 
the instantaneous voltages
$\varphi_t$  for different successive times. From top to bottom
panels we show respectively the first, second and third resonance.
One can see that the voltages follows very well the eigenvectors.

\subsection{Projection on the normal modes}

The normal modes that we have exhibited can be used to project the 
solution. This enables to analyze the dynamical behavior in a
simple way.
The eigenvectors
solutions of the problem (\ref{varepsi_i})
are orthonormal for the inner product defined above. 
We will assume that
all solutions $\phi$ belong to the vector space generated by the 
eigenvectors $\varphi_i$. This is difficult to justify theoretically
but we will come back to this using the numerical results in the 
last section.
Then the solution $\psi$ of equation (\ref{delpsi_phiv})
$\phi$ can be written as
\begin{equation}\label{vctr_spc}
\psi(x;t)=\sum_{i=0}^{+\infty}\beta_i(t)\varphi_i(x).
\end{equation}
We assume the uniform convergence of the series (\ref{vctr_spc}) 
so that we can permute integrals and sums.

We will present the calculations for a single junction for
simplicity. The results can be generalized to arrays with
multiple junctions.
We replace $\psi$ in eq.(\ref{delpsi_phiv}) for one junction,
\begin{eqnarray}
\sum_{i=1}^{+\infty}\left(\beta_i^{\prime \prime}\varphi_i -
\beta_i\varphi_i^{\prime \prime}\right) + \nonumber \\
[-1.5ex] \label{dlt1D_vct} \\ [-1.5ex]
d_1 \delta(x-a_1)\left( \sum_{i=1}^{+\infty}
\left(\kappa \beta_i^{\prime \prime}
\varphi_i + \alpha  \beta_i^{\prime}\varphi_i\right)+
\sin \left(\psi + \Psi_1 \right)-
\frac{\gamma }{d_1 }
\right)&=&0, \nonumber
\end{eqnarray}
We multiply eq.(\ref{dlt1D_vct}) by $\varphi_j(x)$
and integrate it on its domain
\begin{eqnarray}
\sum_{i=1}^{+\infty}\beta_i^{\prime \prime}
\underbrace{\left(\int_0^l \varphi_i \varphi_j dx +
\kappa_1 \varphi_i(a_1)\varphi_j(a_1) \right)}_{=0~{\rm if}~i\neq j,~
=1~{\rm else}}
- \beta_i \int_0^l \varphi_i^{\prime \prime} \varphi_j dx
+d_1 \alpha\beta_i^{\prime}\varphi_i(a_1)\varphi_j(a_1)
 && \nonumber \\
\label{proj1} 
+ d_1 \sin \left(\psi(a_1) + \Psi_1 \right) \varphi_j(a_1)
-\gamma \varphi_j(a_1)
=0. && 
\end{eqnarray}
and from eq.(\ref{eigvp}) we know that,
$$\varphi_i^{\prime \prime}=-\left(1+
\kappa_1 \delta(x-a_1)\right)\omega_i^2 \varphi_i.$$
Eq.(\ref{proj1}) becomes
\begin{eqnarray}
\beta_j^{\prime \prime}+ \sum_{i=1}^{+\infty} 
\beta_i \omega_i^2 \underbrace{ \int_0^l (1+\kappa_1 \delta(x-a_1))
\varphi_i \varphi_j dx}_{=0~{\rm if}~i\neq j,~ =1~{\rm else}}
+\frac{d_1 \alpha}{l}\beta_i^{\prime}\varphi_i(a_1)\varphi_j(a_1)
&& \nonumber \\
+ \frac{d_1 }{l} \sin \left(\psi(a_1)+ \Psi_1\right)
\varphi_j(a_1)-\gamma \varphi_j(a_1) =0. && 
\end{eqnarray}
We obtain the final equation giving the evolution of $\beta_j$
in terms of $\phi$
\begin{equation} \label{rltn2}
\beta_j^{\prime \prime}+ \omega_j^2 \beta_j+
c_j \left( \alpha \phi_t(a_1)
+\sin \left(\phi(a_1) \right) -\frac{\gamma }{d_1}\right)=0,
\end{equation}
where we have returned to the usual field $\phi$ and
where  the coupling coefficient $c_j$ is
\begin{equation}\label{cj}  c_j=d_1 \varphi_j(a_1).\end{equation}
This coefficient does not depend directly on $\kappa$.
Also notice that all equations are coupled by the same term 
\begin{equation}\label{brack} 
F \equiv  \alpha \phi_t(a_1) +\sin (\phi(a_1))-\frac{\gamma }{d_1 }
,\end{equation}
where the coefficient $c_j$ regulates the forcing for each mode.
When $n$ junctions are present in the device, the modal 
equations can be generalized to
\begin{equation} \label{bjt}
\beta_j^{\prime \prime}+ \omega_j^2 \beta_j+
\sum_{k=1}^n c_j^k \left( \alpha \phi_t(a_k)
+\sin \left(\phi(a_k) \right) -\frac{\gamma }
{\sum_{l=1}^n d_l}\right)=0,
\end{equation}
where the generalized coupling coefficient $c_j^k$ is
\begin{equation}\label{cjk}  c_j^k=d_k \varphi_j(a_k).\end{equation}

\section{Numerical analysis of the IV curves}

\begin{figure}
\centerline{\epsfig{file=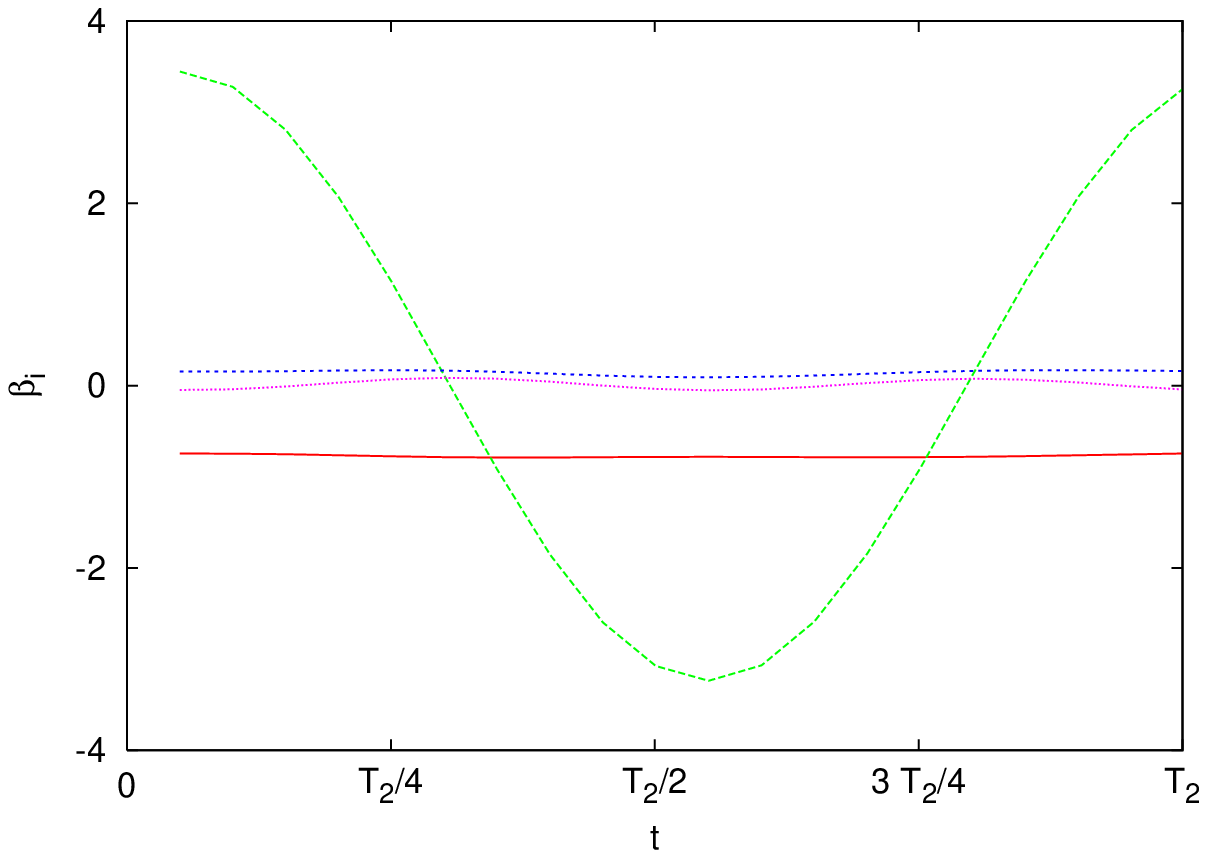,width=0.5\linewidth,angle=0}
\epsfig{file=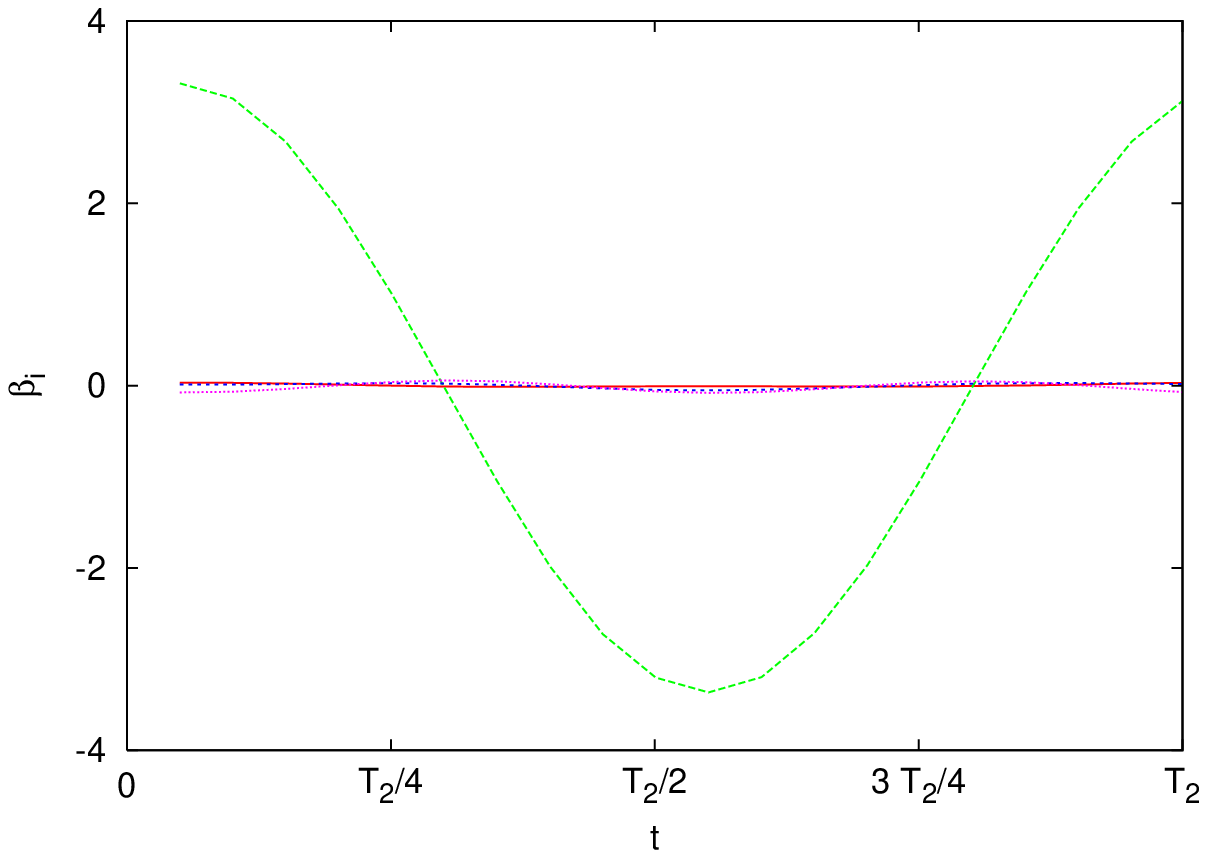,width=0.5\linewidth,angle=0}}
\caption{Plot of the amplitudes $\beta_i(t),~i=1-4$ obtained by projecting the
numerical solution for $\kappa=1$ at the top of the second resonance 
onto the normal mode $\varphi_i$. The modes are $\beta_1$ continuous line,
(red online), $\beta_2$ dashed line (green online), $\beta_3$ short dash (blue
online) and $\beta_4$ dotted line (pink online). In the left panel the
projection is done directly on the solution $\phi(x,t)$. In the right panel
we have subtracted the high voltage solution $\phi_v(x)$ for the given
current $\gamma=1.36$. The time unit is the period $T_2 = 2 \pi / \omega_2 
\approx 3.25$.
}
\label{f5b}
\end{figure}

\begin{figure}
\centerline{\epsfig{file=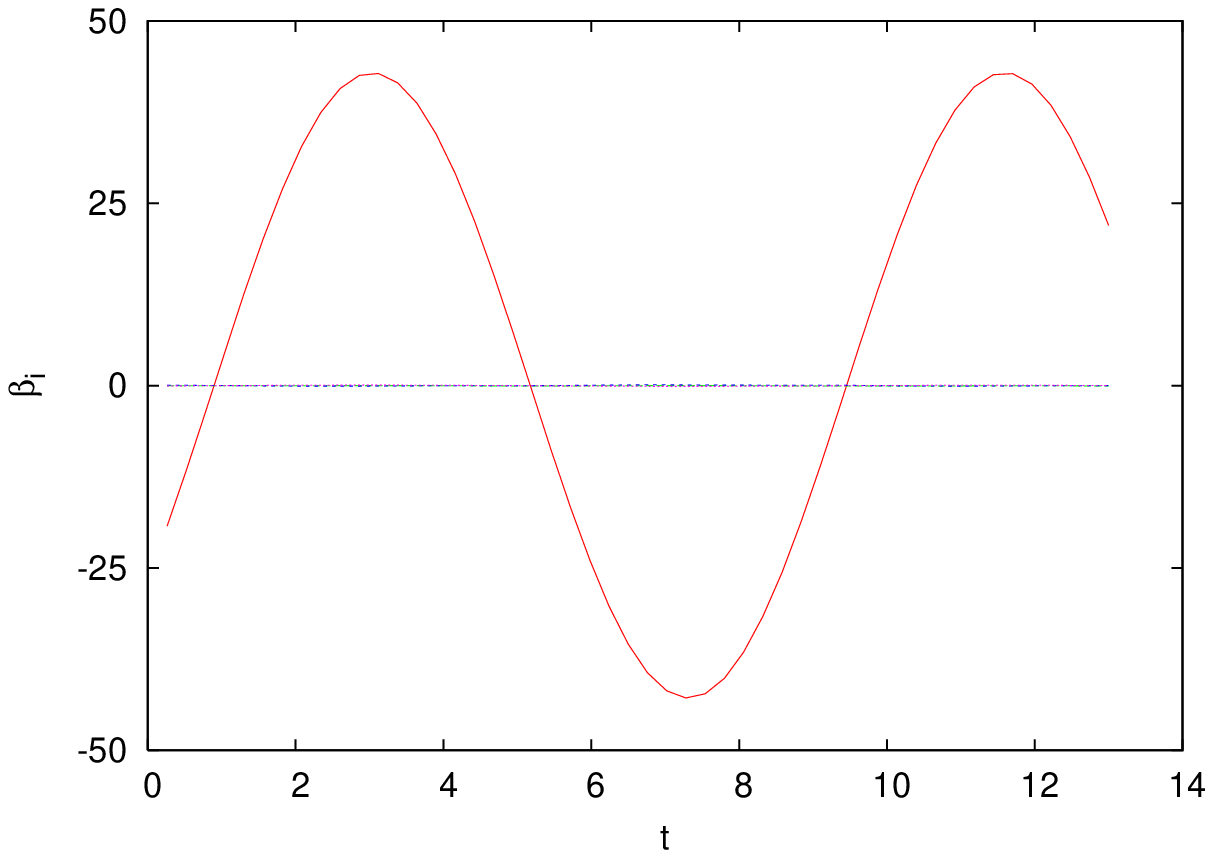,width =0.5\linewidth,angle=0}
\epsfig{file=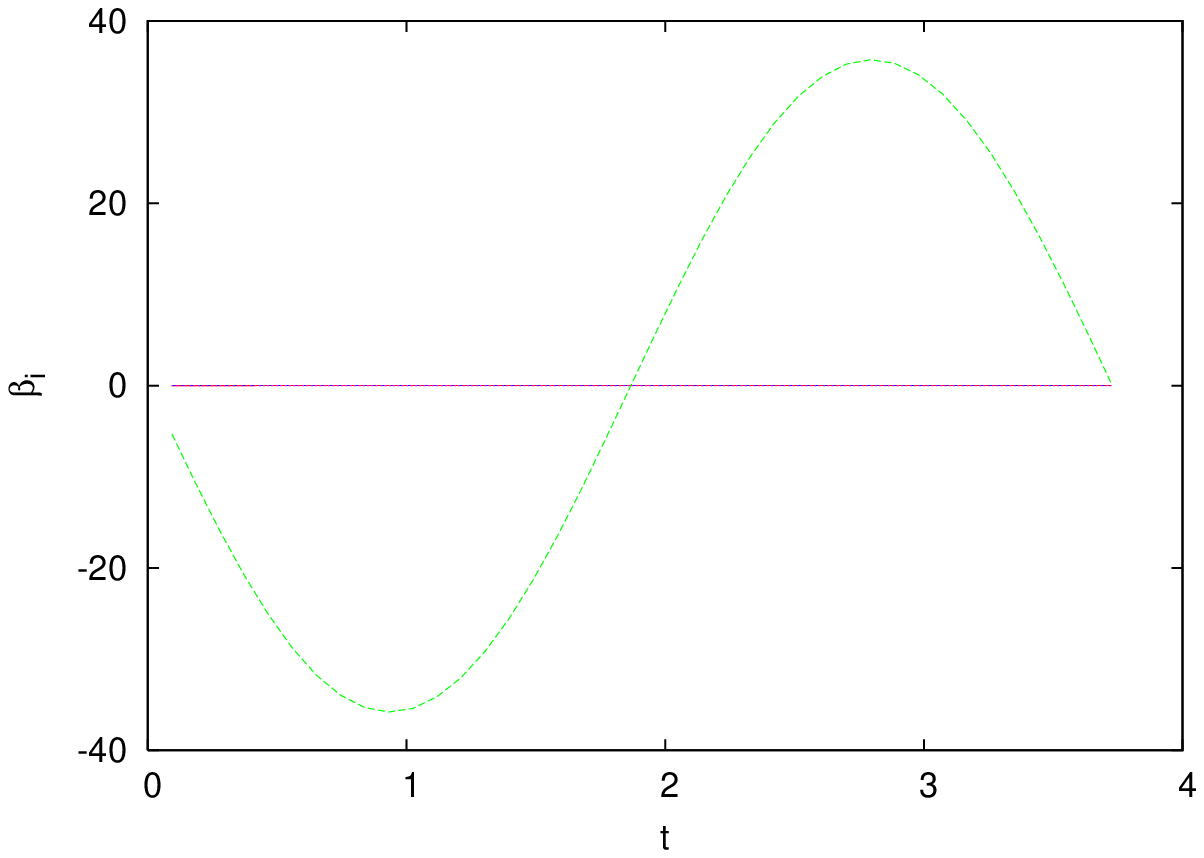,width =0.5\linewidth,angle=0}}
\centerline{\epsfig{file=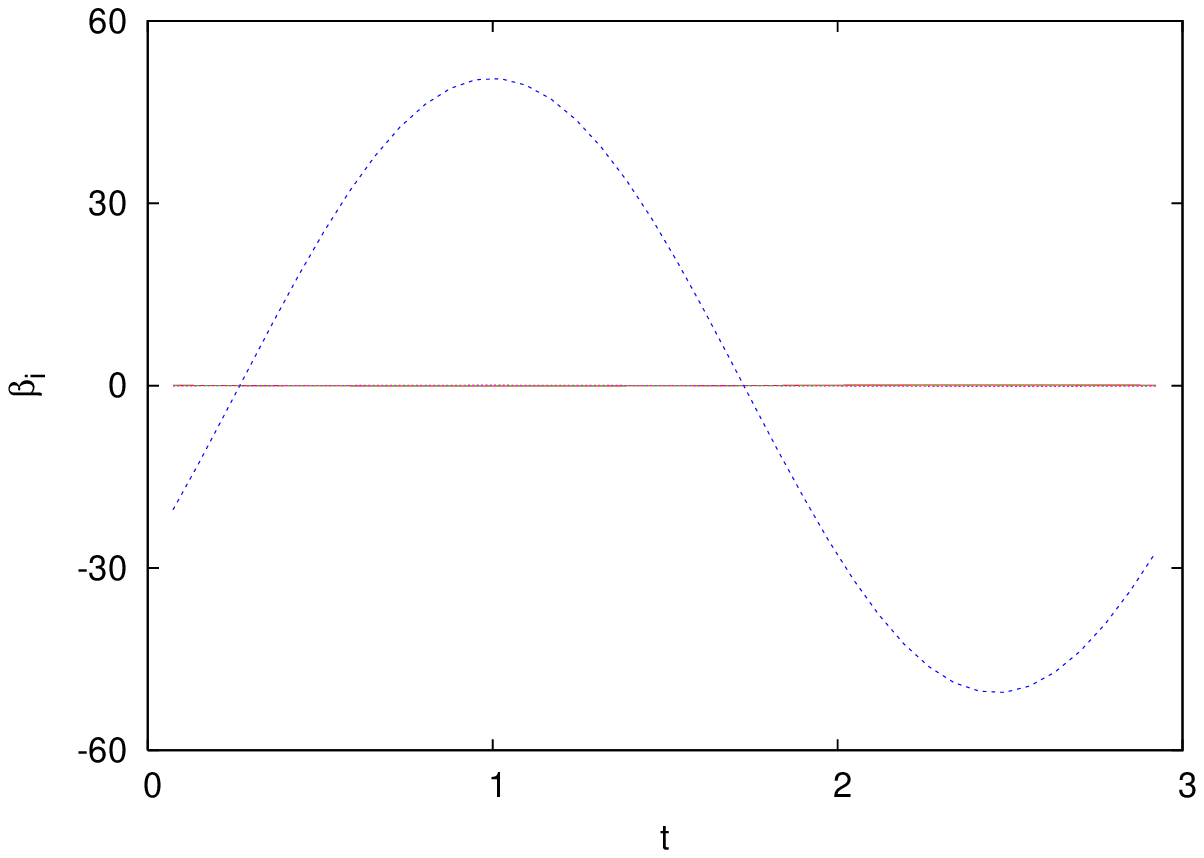,width =0.5\linewidth,angle=0}}
\caption{Plot of the amplitudes $\beta_i(t),~i=1-3$ obtained by projecting the
numerical solution at the top of the second resonance
onto the normal mode $\varphi_i$ for $\kappa=30$. The dominant mode is 
resp. $i=1,2,3$ for resp. the 1st resonance (top panel), the second resonance
(middle panel) and the third resonance (bottom panel).
}
\label{f5c}
\end{figure}

To analyze the mechanism leading to a resonance in the IV curves 
shown in Fig. \ref{f4} we project the numerical
solution onto the normal modes that we defined in section 4. Following
the definition of the inner product (\ref{scl}) we have
$$\beta_i= \left <\phi | \phi_i\right> 
\equiv \int_0^l \phi \phi_i dx + \kappa d_1 
\phi(a_1) \phi_i(a_1) .$$
The integral on the right hand side is computed using the trapeze
method. Fig. \ref{f5b} shows a plot of the amplitudes $\beta_i,~i=1-4$
for the second resonance with $\kappa=1$ (middle panel for
Fig. \ref{f4}). Clearly $\beta_2$ is dominant.
In the left panel we did not subtract the high voltage solution
$\phi_v(x)$ (\ref{phiv0}-\ref{phiv12}) so that the other modes 
appear as parasites. If the high voltage solution is taken out then
we have a clear dominance of $\beta_2$, all the other modes being
close to 0. This shows that we have to a 
good approximation
\begin{equation}\label{phi_res}
\phi(x,t) \equiv \psi(x,t) +\phi_v(x) \approx 
\beta_0(t) \varphi_0(x) + \beta_i(t) \varphi_i(x) + \phi_v(x),\end{equation}
where $i=2$ and where we have included the 0 mode that is always present.
We recover the results suggested by the plots of Fig. \ref{f4a}. We have
observed this for all the resonances in the $I-V$ curve. 
For example for $\kappa=30$ (bottom panel of Fig. \ref{f4}) 
we show in Fig. \ref{f5c} the amplitudes
$\beta_i$ from top to bottom for the 1st , 2nd and 3rd 
resonance. Again the dominant amplitudes are from top to bottom $\beta_1$, 
$\beta_2$ and $\beta_3$.

Therefore the solution at the top of the resonances is
given by (\ref{phi_res}) to a good approximation. To analyze how
we reach this state we have projected the solution on the 
normal modes for increasing values of the current $\gamma$ all the
way to the top of the resonance. The calculations were done over
a long time interval (about 200 periods). Projecting the solution
$\phi$ we observed a drift in the amplitudes $\beta_i$ due
to the rapid increase of $\phi$ and the finite precision
of the evaluation of the integrals. To avoid these technical problems
we have projected the time derivative $\phi_t$. The qualitative
conclusions are the same as for $\phi$ except that we will look at
${\beta_i}_t$. Fig. \ref{f5c} shows three amplitudes as a function of time
for $\gamma=1.12,~1.3,~1.42$ and $1.56$ and a voltage $V \approx \omega_3$
near the 3rd resonance for $\kappa=1$. Only three periods
$T_3= 2\pi / \omega_3$ have been represented for clarity, the rest
of the time evolution is the same. In the top left panel 
for $\gamma=1.12$, the amplitude of the mode 3 is about 1 with small 
components in the modes 2 and 1. When the current is increased the amplitude
of the 3rd mode increases and becomes periodic of period $T_3$. 
The other modes tend rapidly to 0.
\begin{figure}
\centerline{
\epsfig{file=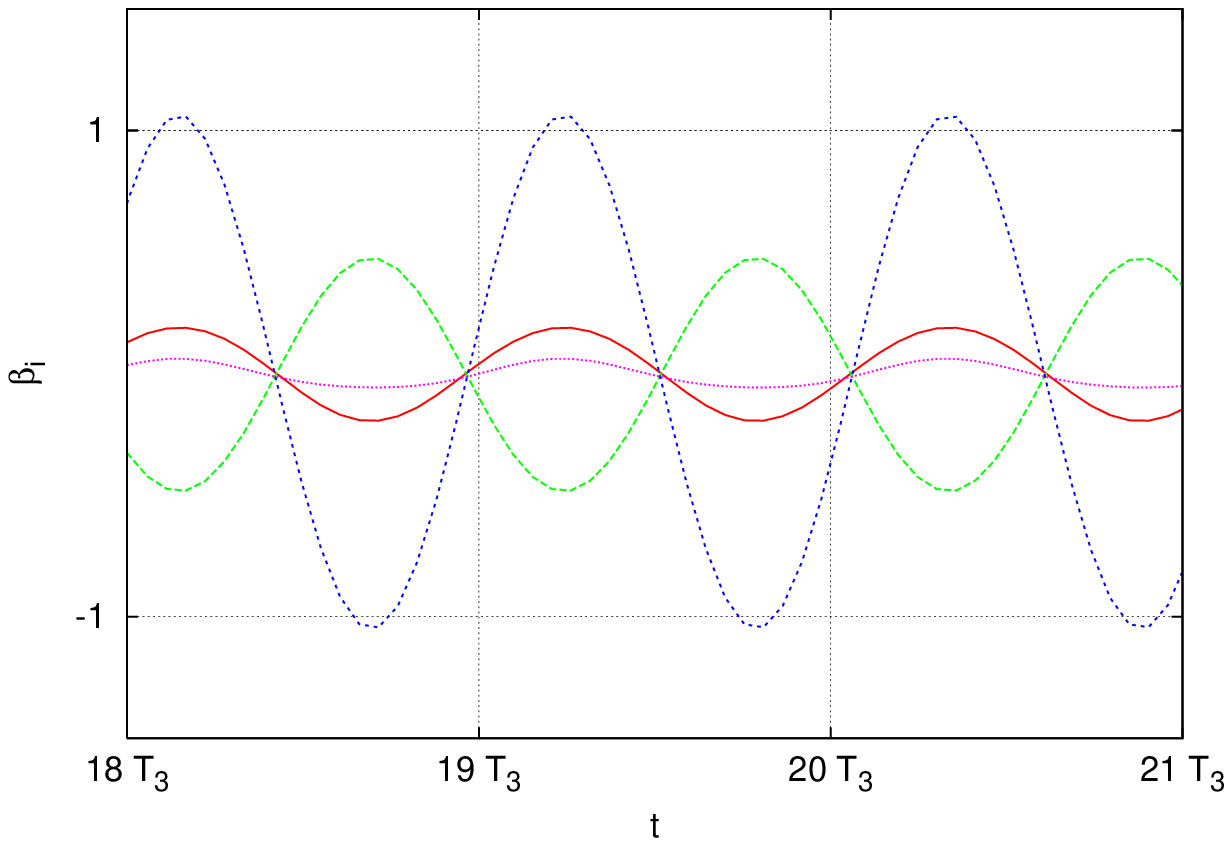,width=0.5\linewidth,angle=0}
\epsfig{file=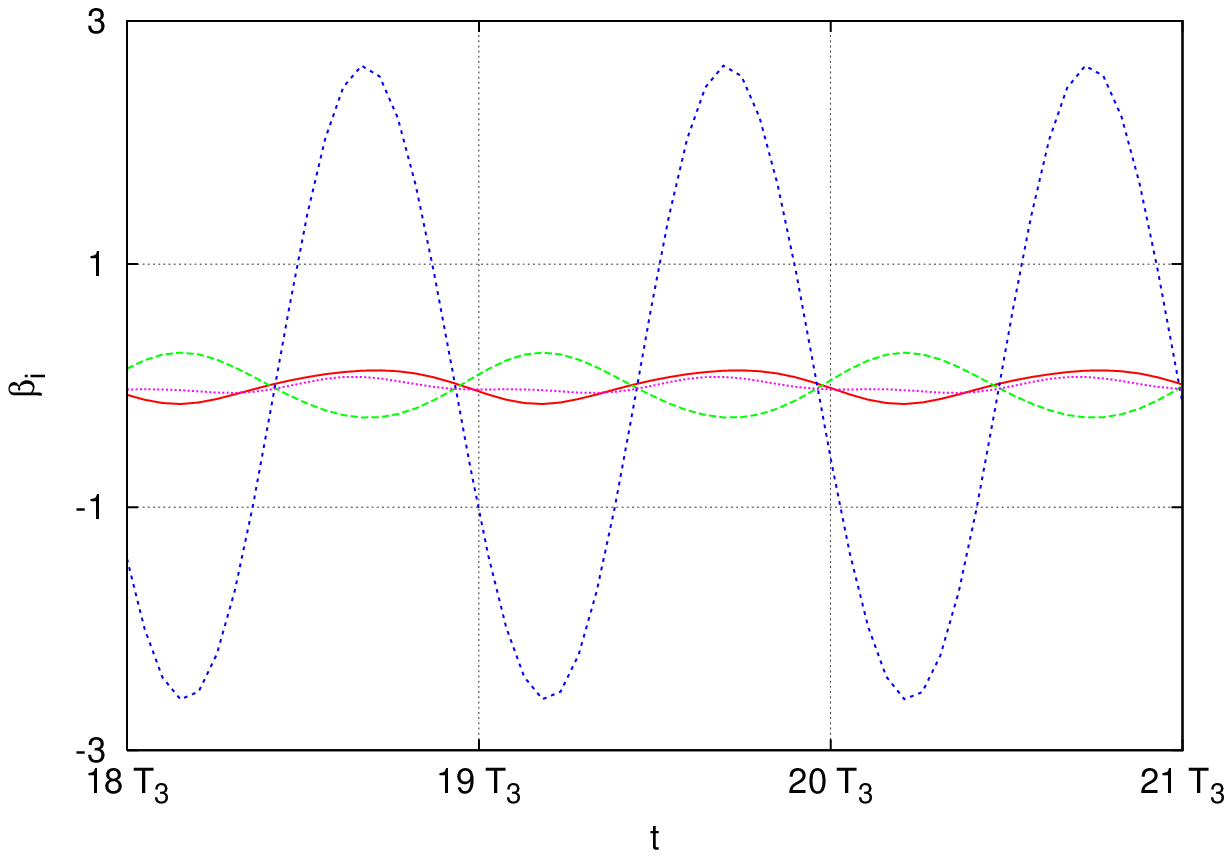,width=0.5\linewidth,angle=0}}
\centerline{
\epsfig{file=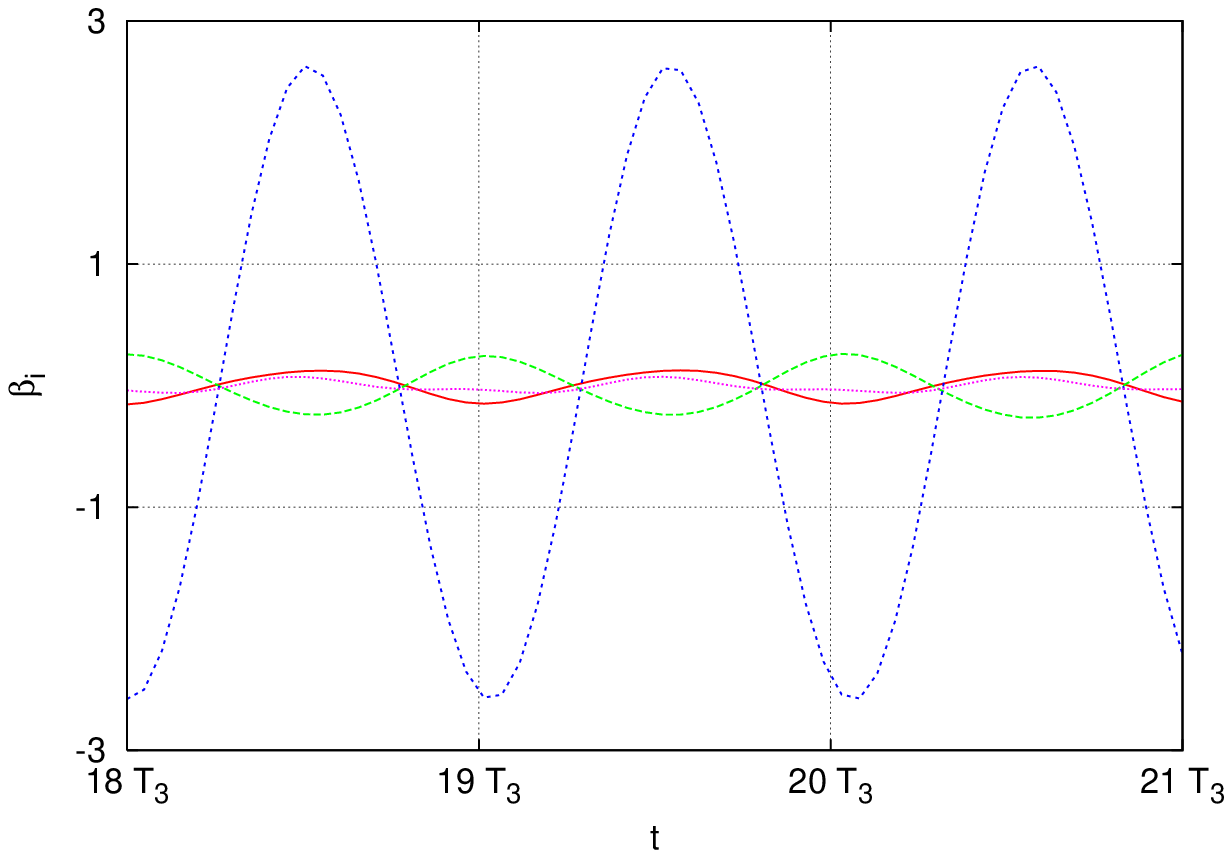,width=0.5\linewidth,angle=0}
\epsfig{file=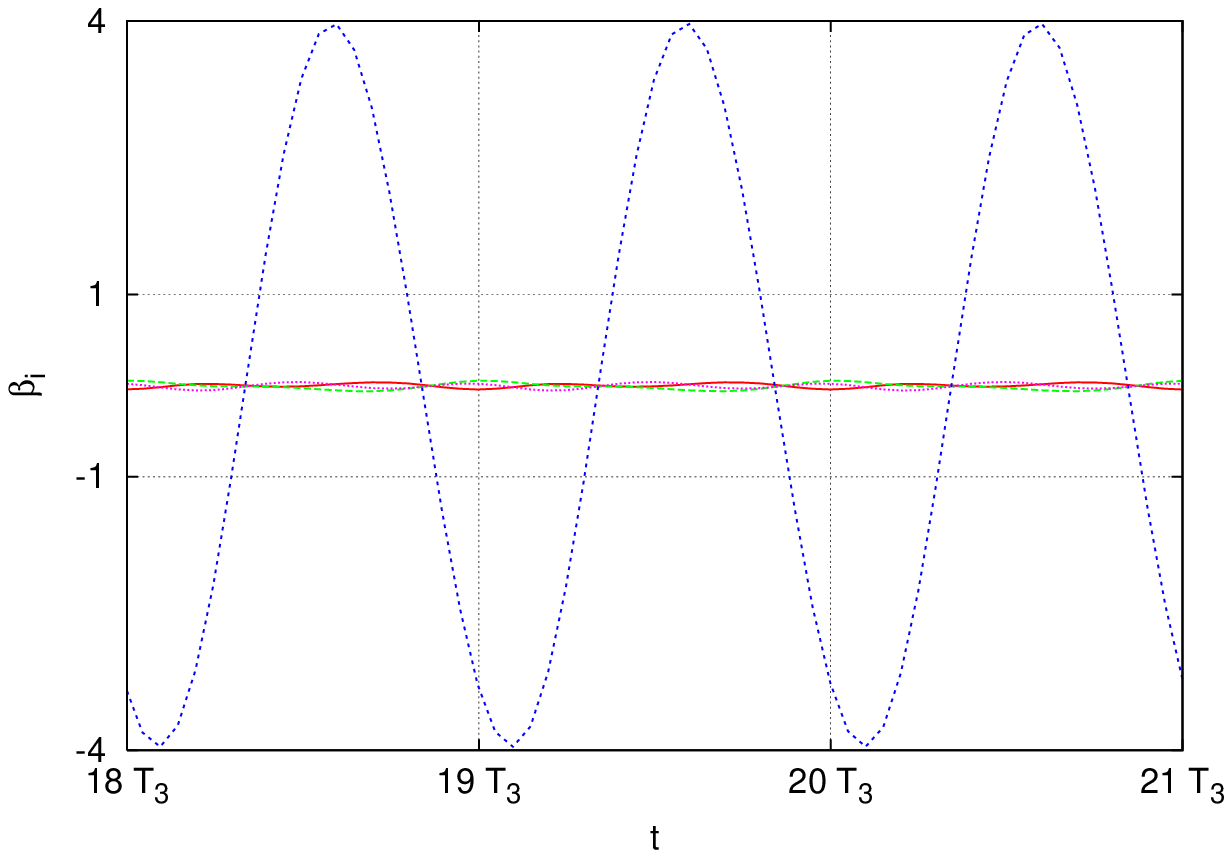,width=0.5\linewidth,angle=0}}
\caption{Plot of the amplitudes of $\phi_t(x,t)$ as a
function of time for four different values of the current, from top left to bottom
right $\gamma = 1.12, 1.3, 1.42, 1.56$, 
on the third resonance $V=\omega_3$ for $\kappa=1$. The index of the modes are
$i=1 $ continuous (red online), $i=2$ dashed line (green online),
$i=3$ short dash (blue online) and $i=4$ dotted
line (pink online). The time interval is the period $T_3=2\pi/\omega_3
\approx 2.58$.
}
\label{f5d}
\end{figure}

\begin{figure}
\centerline{\epsfig{file=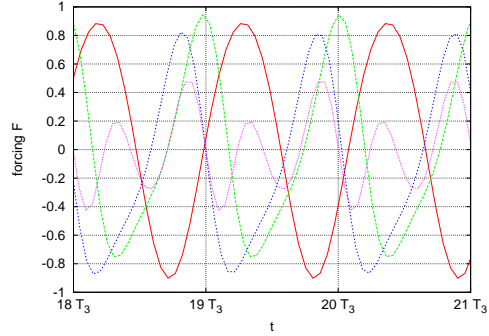,width=0.5\linewidth,angle=0}}
\caption{Plot of the forcing term $F $ from (\ref{brack}) as a 
function of time for four different values of the current on the third 
resonance $V=\omega_3$ for $\kappa=1$. The values are
$\gamma=1.12$ continuous (red online), $\gamma=1.3$ dashed line (green online),
$\gamma=1.42$ short dash (blue online) and $\gamma=1.56$ dotted
line (pink online). The time interval is the period $T_3=2\pi/\omega_3
\approx 2.58$.
}
\label{f5e}
\end{figure}

It is instructive to compute numerically the forcing term $F$
as one progresses up the resonance. Fig. \ref{f5e} shows $F(t)$
for three periods $T_3$ for the four values of $\gamma$ analyzed
in Fig. \ref{f5d}. The amplitude of $F$ decreases for increasing
$\gamma$ and $F$ becomes periodic of period $T_3$. This explains
why we obtain the correct resonant modes using the 
spectral problem (\ref{eigvp}). Note that when $\kappa=0$ the
forcing term $F$ of the amplitude equations tends to 0 when one
gets close to the top of the resonances. Then one can solve the
differential equation for $\phi(a_1;t)$
\begin{equation}\label{odephi}
\alpha \phi^{\prime}(a_1;t) +\sin \left(\phi(a_1;t)\right)
-\frac{\gamma }{d_1 }=0,
\end{equation}
and close the system by expanding this solution using the standard
Fourier modes\cite{cl05}. This is not the case when $\kappa \neq 0$ as shown
by these numerical results.
To analyze the resonances, we assume as in \cite{kulik} that when in
resonance, the solution  has the spatial structure of the corresponding
eigenmode. 
\begin{equation}\label{reson_anstaz}
\psi(x,t) = \beta_0(t) \varphi_0 + \beta_n(t) \varphi_n(x) ,\end{equation}
where the first term on the right corresponds to the
zero mode.
The evolution of $\beta_0,~\beta_n$ is then given by equation (\ref{rltn2})
\begin{eqnarray}\label{betat}
\beta_n^{\prime \prime}+ \omega_n^2 \beta_n+ c_n F =0, \\
\beta_0^{\prime \prime}+ c_0 F =0, 
\end{eqnarray}
where the forcing term is
\begin{equation}\label{force}
F = \alpha \beta_n^{\prime} \varphi_n 
+ \alpha \beta_0^{\prime} \varphi_0   +
\sin \left( \beta_n \varphi_n  + \beta_0 \varphi_0 \right)  
-\frac{\gamma }{d_1 }   
\end{equation}
To understand the specific shape of the resonances, the fact that
we cannot obtain in the IV curve the right part of the resonance 
curve, one could carry out a bifurcation analysis similar to the
one of \cite{bfmp94}. However this is out of the scope of this
article.

\begin{figure}
\centerline{\epsfig{file=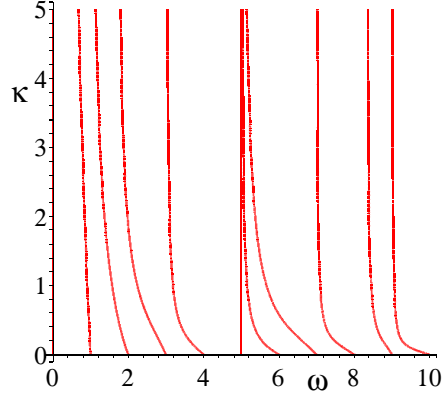,width=0.5\linewidth,angle=0}}
\caption{Plot of the frequency $\omega$ vs. $\kappa_i$ for a
device with two junctions where
$a_1=0.3 \pi,~a_2=0.5\pi,~ d_1=d_2=0.1$.}
\label{f5}
\end{figure}

\begin{figure}
\centerline{ \epsfig{file=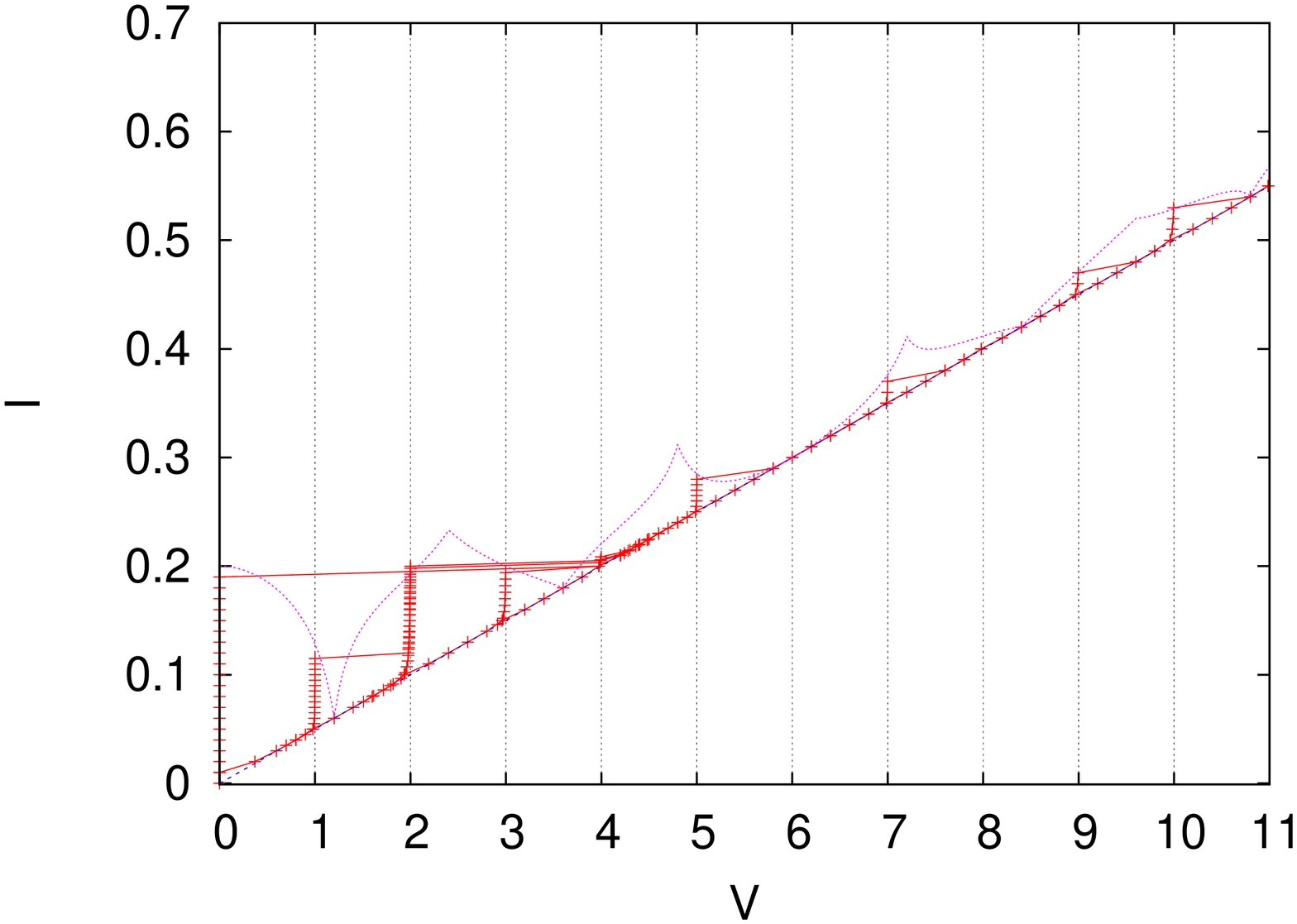,height=0.5\linewidth,angle=270} 
\epsfig{file=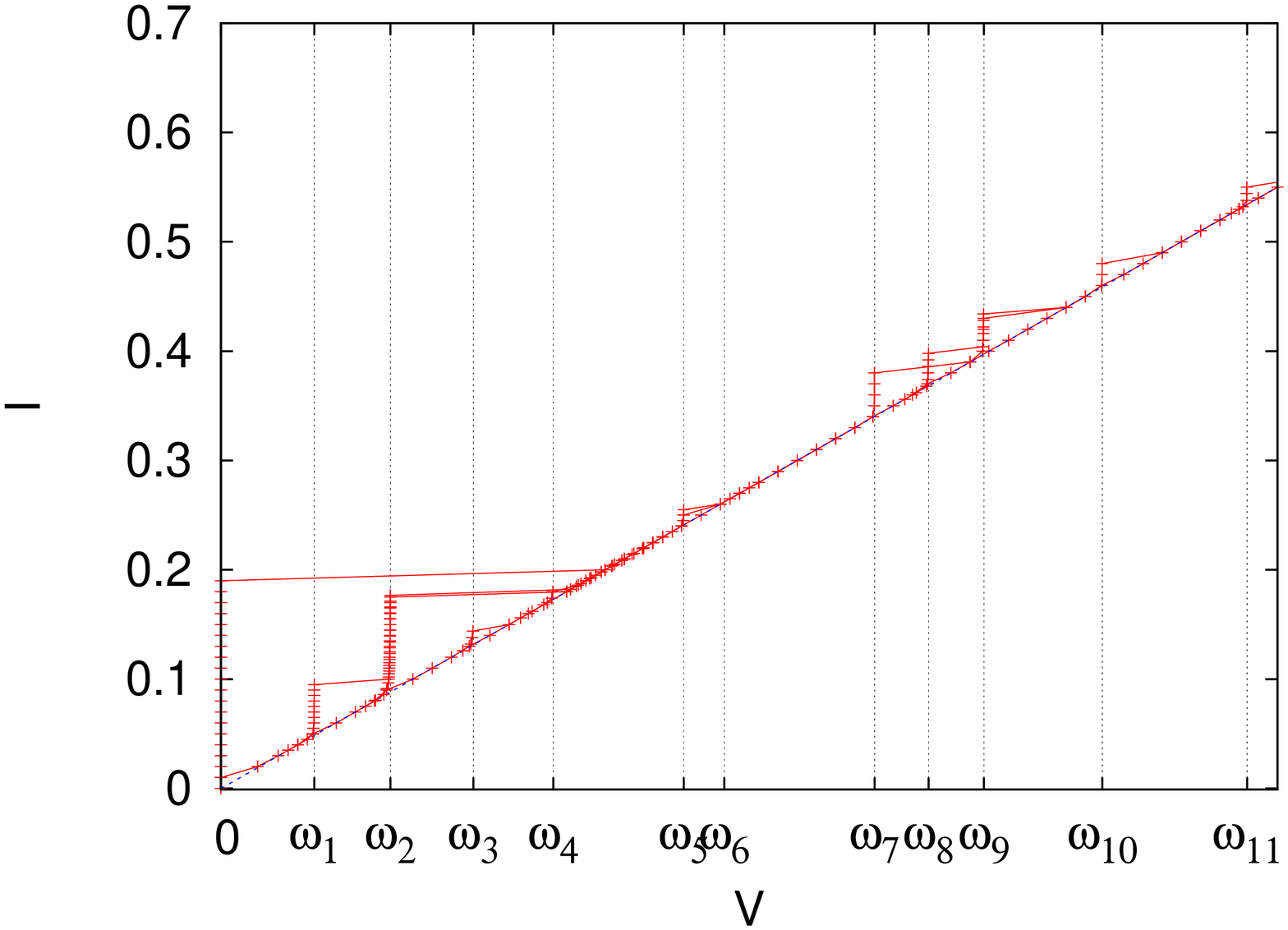,height=0.5\linewidth,angle=270} }
\centerline{\epsfig{file=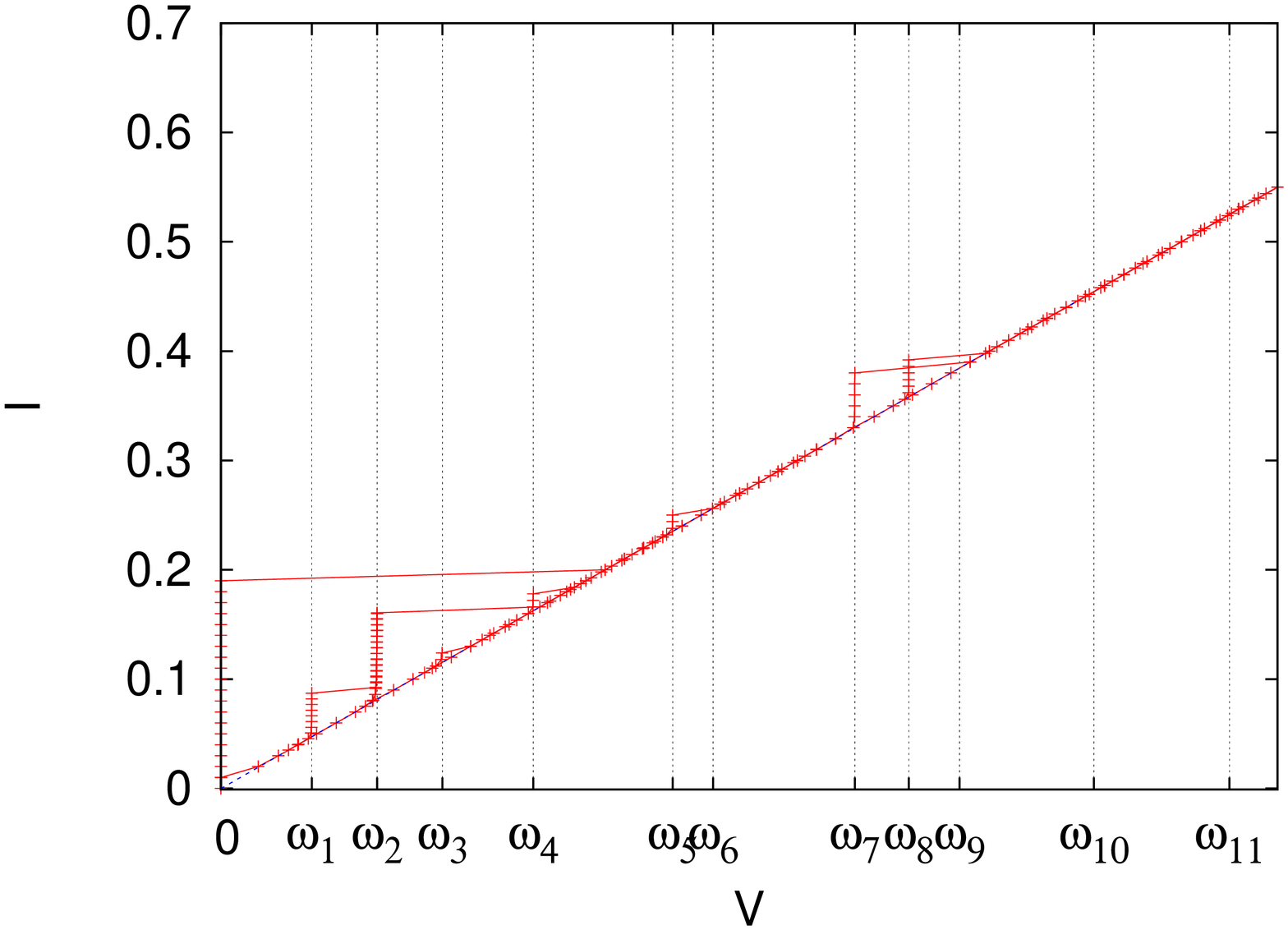,height=0.5\linewidth,angle=270} }
\caption{IV curves for the two junction circuit studied in Fig. \ref{f5}. 
The parameters are
$l=\pi, d_1=d_2=0.1,~\alpha=0.25,~\kappa=0$. From top left to bottom $\kappa =0,4$
and 8. In the top panel the envelope function (see text)
is plotted as a dashed line (pink online).}
\label{f6}
\end{figure}

\begin{figure}
\begin{minipage}{0.49\linewidth}
\epsfig{file=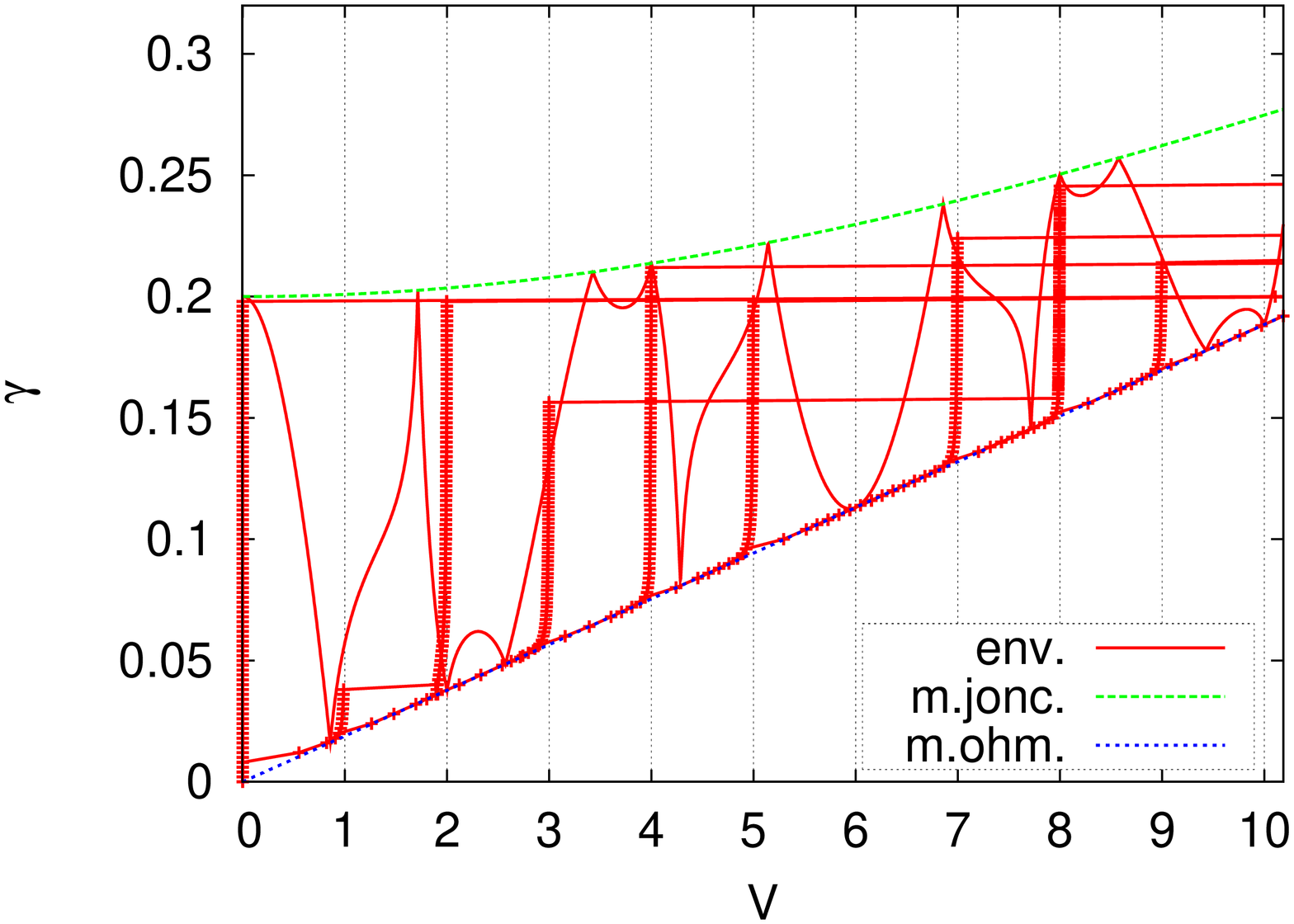,height=\linewidth,angle=270}
\end{minipage}
\begin{minipage}{0.49\linewidth}
\epsfig{file=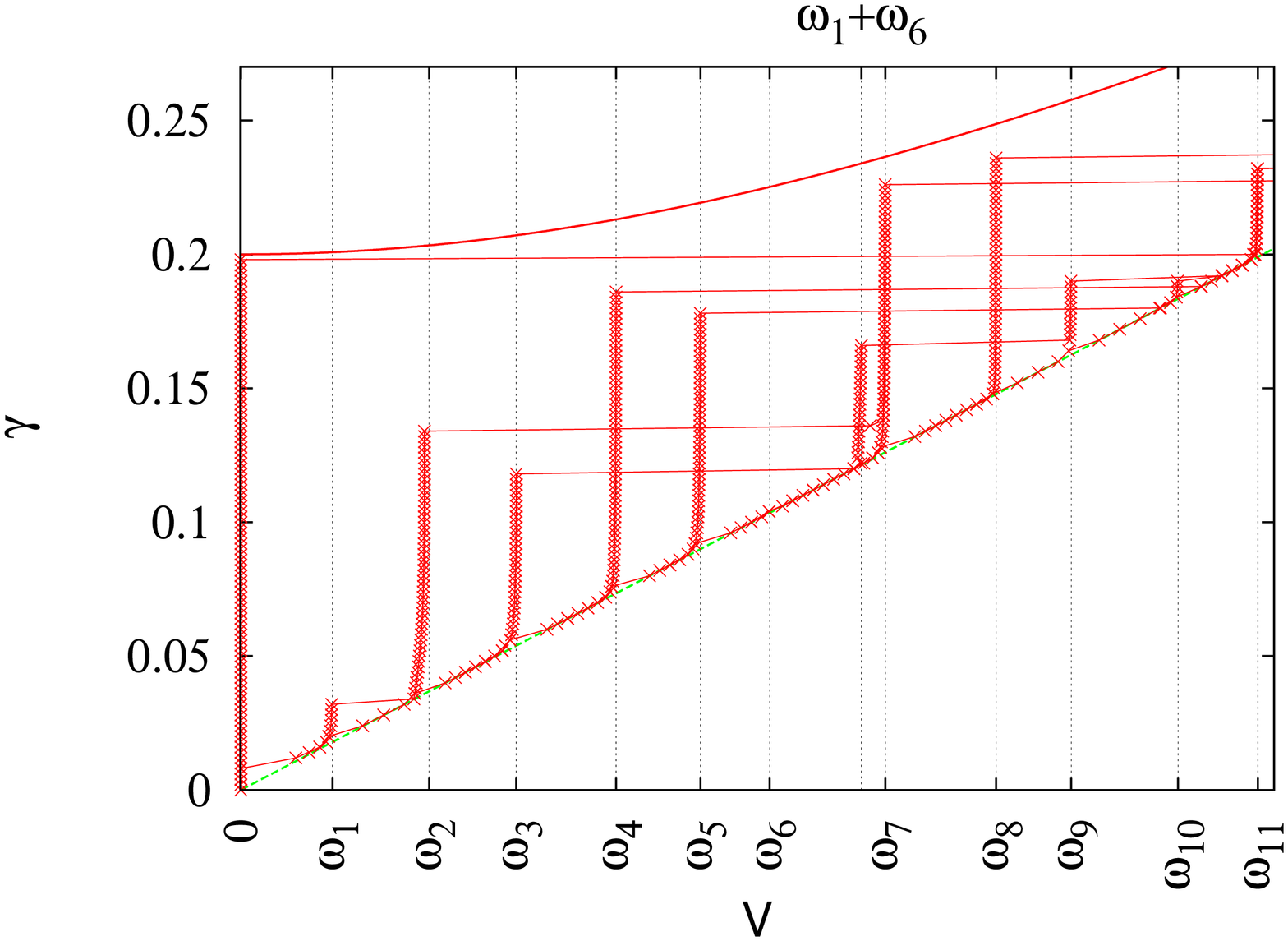,height=\linewidth,angle=270}
\end{minipage}
\begin{minipage}{0.49\linewidth}
\epsfig{file=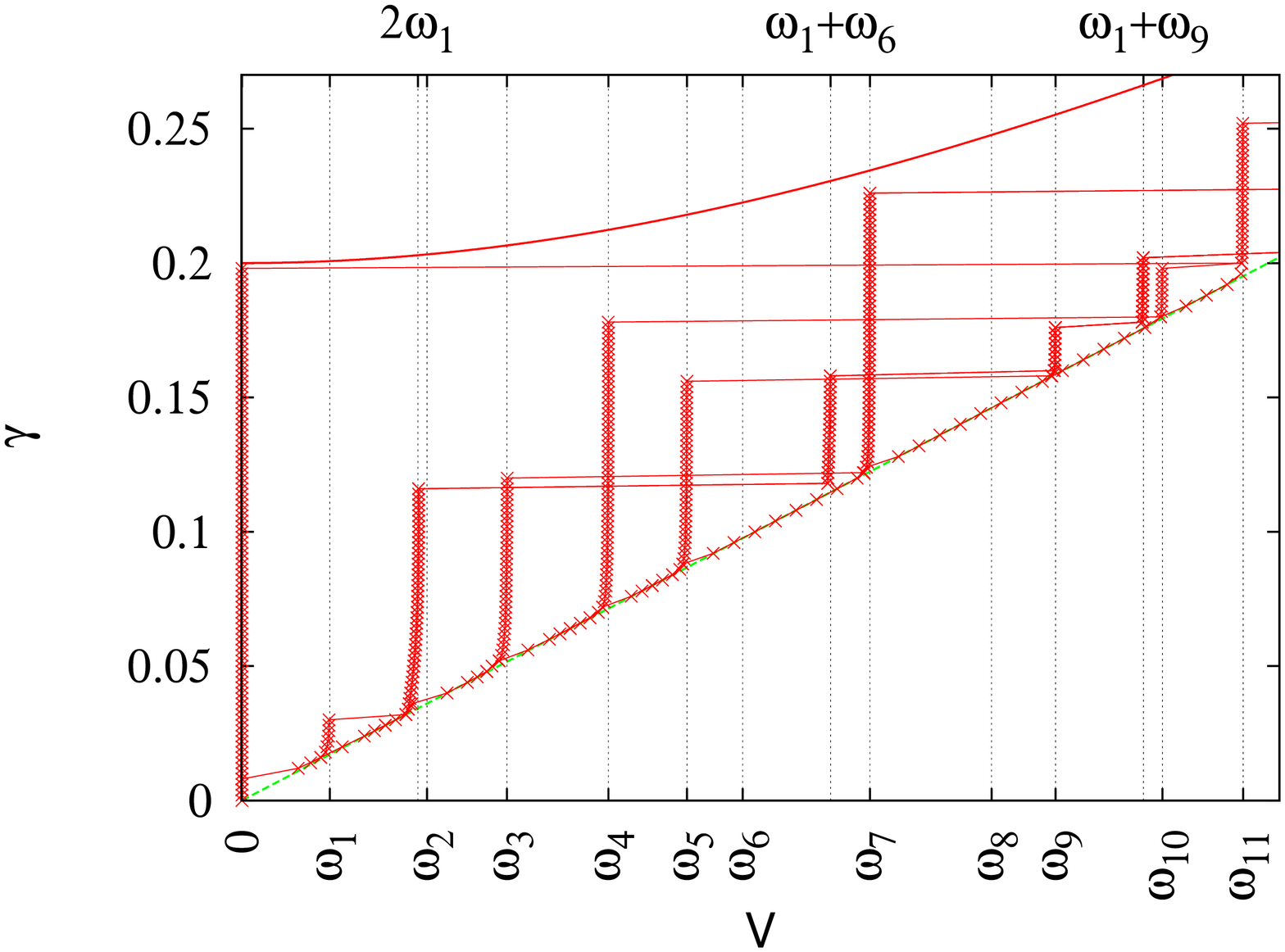,height=\linewidth,angle=270}
\end{minipage}
\begin{minipage}{0.49\linewidth}
\epsfig{file=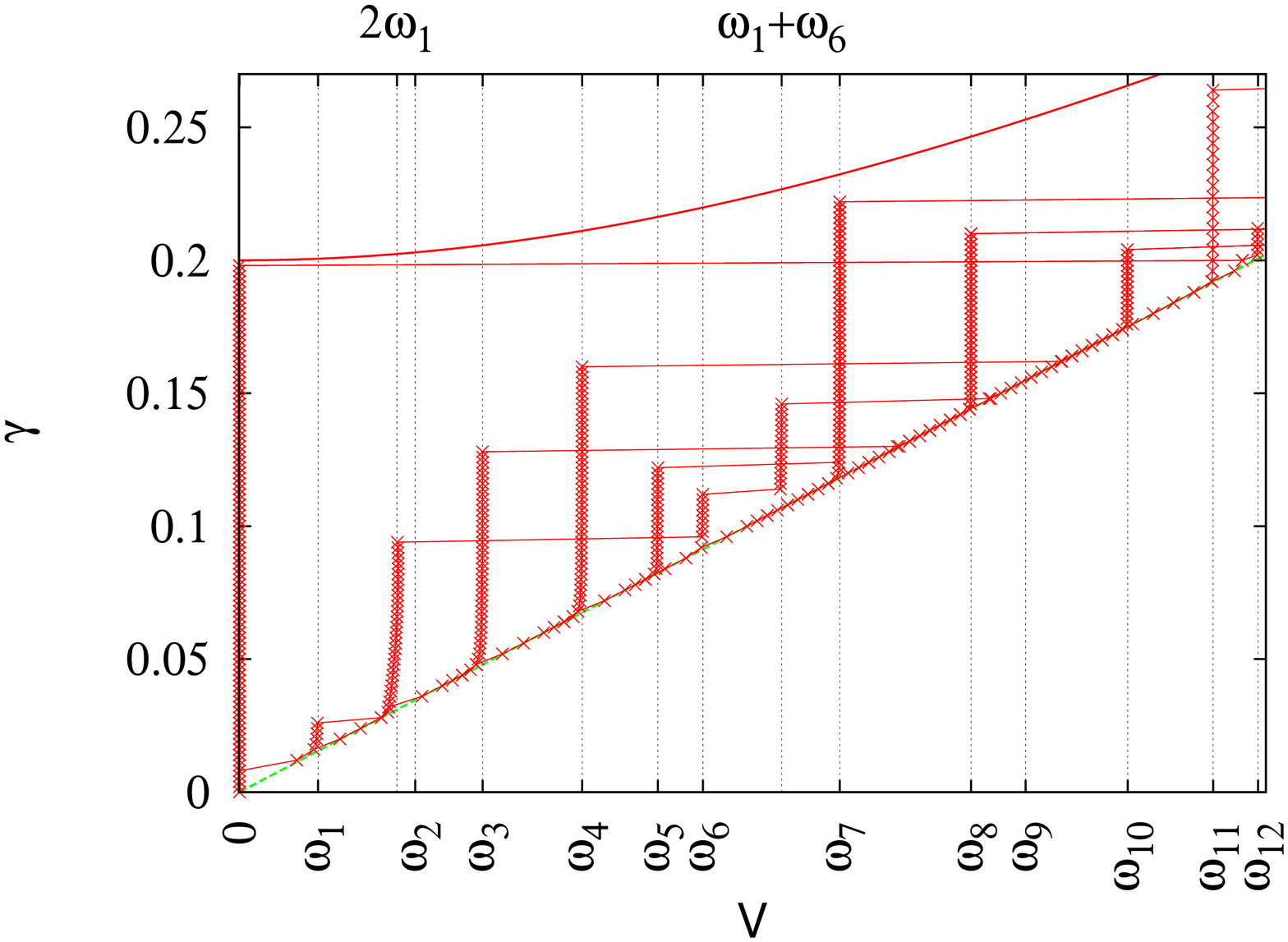,height=\linewidth,angle=270}
\end{minipage}
\label{f10}
\caption{\it $I-V$ curves of a two junction device, $a_1=10/3$, $a_2=25/3$,
$l=10$, $d_i=0.1$, $\alpha=0.3$ and different values of
$\kappa$. Top panels, $\kappa = 0$ (left) and $\kappa = 5$ (right).
Bottom panels $\kappa = 10$ (left) and $\kappa = 22 $ (right).}
\end{figure}

The situation is more complex when there are more junctions in the
device. As an example we consider a two junction device
with $l=\pi, ~a_1=0.3 \pi,~a_2=0.5\pi,~ d_1=d_2=0.1$. The eigenmodes
$\omega_n$ are plotted in Fig. \ref{f5} as a function of $\kappa$ and
one can see them shift from integer values even for small $\kappa$.
In Fig. \ref{f6} we plot the $I-V$ curves obtained for this device
for $\kappa =0,~4$ and 8 with $d_1=0.1$ so that 
$\kappa_1 = d_1 \kappa = 0,~0.4$ and 0.8. The resonances observed correspond
to the eigenfrequencies obtained.
For $\kappa=0$ we have explained the height of the resonances using 
an approximate theory \cite{cl07} based on the amplitude 
of oscillation of $\phi_t(a_j)$ for each junction. This is the
envelope function plotted in dashed line in the top left panel of
Fig. \ref{f6}.  When $\kappa \geq 0$ it is more difficult. We think
that this amplitude of oscillation, which can be obtain by the eigenvectors
does not determine completely the height of the resonances. 
When the length of the device is larger more resonances can be accommodated
in the $I-V$ curve. An interesting effect we found is that
for large values of $\kappa$ the system locks to linear combinations
of the eigenfrequencies. Such an example is shown in Fig. \ref{f10}
for a two junction device in a microstrip of length $l=10$.
For $\kappa \ge 5$ there appears in the $I-V$ curve a resonance
for $\omega_1+\omega_6$. For $\kappa=10$ (bottom left panel) we see
in addition 
resonances for $2 \omega_1, ~\omega_1+\omega_9$. This is typical
of a linear system. When we observe numerically 
the evolution of $\phi_t$ at this resonance we see the two eigenvectors 
1 and 9. But we cannot explain this linear behavior: how the system
can sum two non linear solution?

\subsection{Conclusion}
To summarize we have analyzed a long wave model describing a parallel
array of Josephson junctions. We defined an appropriate spectral
problem whose spectrum gives the resonances of any array with
junctions of arbitrary (small) sizes and positions. This task
would not have been possible without the complexity reduction
provided by asymptotic analysis.

The adapted spectral problem leads to an inner product so that
it becomes possible to project the dynamics of the system and
describe arrays with more junctions.

It may now be possible to solve the inverse problem of finding the
device yielding a given I-V curve. Another open question is the
study of the amplitude equations (\ref{betat}) to analyze the stability
of the resonances.

{\bf Acknowledgements}

J.G.C. and L. L. thank Faouzi Boussaha and Morvan Salez for helpful discussions
and for their experimental results.
The computations were done at the Centre de Ressources
Informatiques de Haute-Normandie (CRIHAN).

\section{Appendix\label{appendix_num}}
The basis of the method is to discretize the spatial part of the
operator and keep the temporal part as such. We thereby transform
the partial differential equation into a system of ordinary
differential equations. This method allows to increase the
precision of the approximation in time and space independently and
easily. In our case the operator is a distribution so that the
natural way to give it meaning is to integrate it over a volume.
We therefore choose as space discretisation the finite volume
approximation where the operator is integrated over reference volumes.
The value of the function is assumed constant in each volume.
As solver for the system of differential equations, we use
the Runge-Kutta method of order 4-5 introduced by Dormand and Prince
implemented as the Fortran code DOPRI5 by Hairer and Norsett \cite{hairer}
which enables to control the local error by varying the time-step.

We first transform (..) into a system of first order partial
differential equations
We write $\psi(x,t)=\phi_t(x,t)$.
\begin{equation} \label{sysdiscret}
\left\{  \begin{array}{r c l}
\psi(x,t) & = & \phi_t(x,t) \\
\psi_t(x,t) & = & \phi_{xx}(x,t)-\delta(x-a) 
(\kappa \psi_t(x,t) + \alpha \psi(x,t)) +d_1 \sin(\phi(x,t)))
+ \nu \gamma / l
\end{array} \right.
\end{equation}
with the boundary conditions :
$\phi_x|_{l \over 2}=H-(1-\nu) \gamma /2$, and
$\phi_x|_{-{l \over 2}}=H+(1-\nu) \gamma /2$.

For simplicity we will describe the implementation of
the finite volume discretisation in the case of a single junction.
We introduce reference
volumes $V_k$ whose centers we call $x_k$, $1\leq k \leq nn$.
The discretisation points are placed such that
the point $x_{ng+1}$ is at the junction, ($x_{ng+1}=a$).
We thus define $x_k$ and $V_k$ using the following identities
$$V_k=\left ]x_k-{h_g \over 2},x_k+{h_g \over 2}
\right[~,~~0<k<ng+1$$
with $(ng+1)h_g=a$
$$V_k=\left ]x_k-{h_d \over 2},x_k+{h_d \over 2}
\right[~,~~ng+1<k<nn+1$$
with $(nn-ng)h_d=l-a$. Finally at the junction, $k=ng+1$
$$V_{k_{ng+1}}=\left ]x_{ng+1}-{h_g \over 2},
x_{ng+1}+{h_d \over 2} \right[.$$
$nn$, $ng$ and $nd$ are respectively the total number of discretisation
points, the number of points to the left of the junction and the number
of points to the right.

For a fixed t, we assume $\phi(x,t)$ to be constant on each volume $V_k$,
so that
$$\int_{x_k-{h \over 2}}^{x_k+{h \over 2}} \phi(x,t) dx =
h \phi(x_k,t)~{\rm,~~with}~h=hg~{\rm or}~h=hd$$
Integrating over $V_k$ yields:

In the linear part of the partial differential
equation : $0<k<nn+1$ and $k \neq ng+1$:
\begin{equation} \label{spacediscr}
\left \{  \begin{array}{r c l}
\psi(x_k,t) & = & \phi_t(x_k,t) \\
\psi_t(x_k,t) & = &
\frac{\phi(x_{k+1},t)-2\phi(x_k,t)+\phi(x_{k-1},t)}{h^2} + j
\end{array} \right.
\end{equation}
with $h=hg$ for $0<k<ng+1$ or $h=hd$ for $k>ng+1$. We recognize the usual
discretisation of the second derivative.

At the junction: $k=ng+1$, we obtain
\begin{eqnarray*}\label{disconti}
\int_{x_{ng+1}-{h_g \over 2}}^{x_{ng+1}+{h_d \over 2}}
\delta(x-a)\left(\kappa \phi_{tt}(x,t)
+ \alpha\phi_t(x,t) + d_1 \sin(\phi(x,t))\right)=\\
d_1 \sin(\phi(x_{ng+1},t)) + \alpha\phi_t(x_{ng+1},t)
\end{eqnarray*}
So that the final system is:
\begin{eqnarray*}
\psi(x_{ng+1},t) & = & \phi_t(x_{ng+1},t) \nonumber\\
\psi_t(x_{ng+1},t) & = & 
\left[ \frac{4}{hg+hd} \left(\frac{\phi(x_{ng+2},t)-
\phi(x_{ng+1},t)}{hg/2} -\frac{\phi(x_{ng+1},t)-\phi(x_{ng},t)}
{hd/2} \right)\right. \\  
 & & \left. -\frac{2}{hg+hd}\left(d_1 \sin(\phi(x_{ng+1},t))
 +\alpha\phi_t(x_{ng+1},t)\right) + j \right] 
\frac{1}{1-d_{k} \kappa}
\end{eqnarray*}

The previous system of ordinary differential equations
is then integrated numerically using the DOPRI5 integrator.

\end{document}